\documentclass[aps,twocolumn,amsmath,amssymb,showpacs,showkeys,floatfix,nofootinbib]{revtex4}
\usepackage{graphicx}
\usepackage{amssymb}
\usepackage{color}
\usepackage{float}
\usepackage{mathrsfs}
\usepackage{epstopdf}
\usepackage{siunitx}
\usepackage{hyperref}
\usepackage[normalem]{ulem}

\newcommand{\e}[1]{\num[exponent-product=\cdot,exponent-base=10]{#1}}

\begin{document}

\title{{Interplay between low-lying isoscalar and isovector dipole modes: \\ a comparative analysis between semi-classical and quantum approaches}}

\author{S. Burrello$^{1}$, M. Colonna$^{1}$,  G. Col\`{o}$^{2,3}$,
D. Lacroix$^{4}$, X. Roca-Maza$^{2,3}$, G. Scamps$^{5,6}$, H. Zheng$^{1,7}$}

\affiliation{$^{1}$ Laboratori Nazionali del Sud, INFN, I-95123 Catania, Italy}
\affiliation{{$^{2}$} Dipartimento di Fisica, Universit{\`a} degli Studi di Milano, via Celoria 16, Milano 20133, Italy}
\affiliation{{$^{3}$} Istituto Nazionale di Fisica Nucleare (INFN), Sezione di Milano, via Celoria 16, Milano 20133, Italy}
\affiliation{$^{4}$ Institut de Physique Nucl\'eaire, IN2P3-CNRS, Universit\'e Paris-Sud, Universit\'e Paris-Saclay, F-91406 Orsay Cedex, France}
\affiliation{{$^{5}$} {Institute of Astronomy and Astrophysics (IAA), Universit\'e libre de Bruxelles (ULB) , CP 226, Boulevard du Triomphe, B-1050 Bruxelles, Belgium}}
\affiliation{{$^{6}$} Center  for  Computational  Sciences, University of Tsukuba, Tsukuba  305-8571, Japan}
\affiliation{{$^{7}$} School of Physics and Information Technology, Shaanxi Normal University, Xi'an 710119, China}

\begin{abstract}
We perform Time Dependent Hartree-Fock (TDHF) calculations to investigate
 the small amplitude dipole response of selected neutron-rich nuclei and Sn isotopes. 
A detailed comparison with the dipole strength predicted by Random-Phase Approximation (RPA) calculations is presented for the first time. TDHF results are
also confronted to Vlasov calculations,
to explore up to which extent a semi-classical picture can explain
the properties of the nuclear response.  
The focus is on the low-energy response, below the Giant Dipole Resonance region, {where different modes} of non negligible strength are identified. We show that the relative weight 
of these excitations evolves with nuclear global features, such as density profile and neutron skin, which in turn reflect important properties of the nuclear effective interaction.
 A thorough analysis of the associated transition densities turns out to be 
quite useful to better characterize the mixed isoscalar(IS)/isovector(IV) nature {of the different modes} and their surface/volume components.   
{In particular, we show that the dipole response in 
the so-called Pygmy Dipole Resonance region  corresponds to isoscalar-like surface oscillations, of larger strength in nuclei with a more diffuse surface. 
The ratio between the IV and IS Energy Weighted Sum Rule fractions exhausted in this region is shown to almost linearly increase with the neutron skin thickness
in Sn isotopes.}

\end{abstract}



\maketitle

\section{Introduction}

The development of collective motion, i.e., a coherent pattern of particles in phase--space, is a fundamental feature of many-body systems. For instance, atomic nuclei are strongly correlated, self-bound many-body systems, which, together with single-particle properties, exhibit a variety of collective phenomena~\cite{Bohr1969, Row2010}. 
More recent examples are provided by Bose-Einstein condensates~\cite{butNAT1999, madPRL2000, marPRL2000} and there are strong experimental and theoretical evidences that similar effects occur in other fermionic systems as well~\cite{oosPRL1999,torPRL2004}.
Collective patterns may emerge out of single-particle motion whenever favored by energy and kinematic conditions. The collective dynamics is often well described at the classical level and it is reflected in the spectral properties of the corresponding quantum many-body system. Hence, the associated spectrum of excitations usually includes states of single-particle and of collective natures, together with mixed forms, with a partial degree of collectivity. A 
{thorough understanding} of the emergence of collective motion from the microscopic point of view is however a true challenge. 

The isovector Giant Dipole Resonance (GDR) in heavy nuclei is a prominent and well known example of collective motion, first measured in photo-absorption experiments~\cite{BalPR1947}. 
The cross-section associated with the electric dipole radiation and the corresponding strength function show between 10-30 MeV - the energy depending on the size of the nucleus as $\approx$ A$^{1/3}$ - a large increase, with a spreading width 
larger than the mean level spacing. This excitation can be described in terms of the classical picture of neutrons and protons moving against each other, resulting in a large response function. Thus the difference between the center-of-mass coordinates of the two spheres appears as the proper collective coordinate in this case. 

{In recent years, there has been a considerable amount of experimental and theoretical studies on dipole excitations {in neutron-rich nuclei}, {and in particular} on the low-energy tail of the isovector GDR, 
the so-called Pygmy Dipole Strength (PDS) observed in the IV dipole response
(often denominated Pygmy Dipole Resonance (PDR))~\cite{PaarPLB2005, SavPRC2011, CrePRL2014}.} 

The PDR has been often interpreted as an exotic mode of excitation due to the motion of the weakly bound neutron excess against an almost inert proton-neutron core~\cite{paaRPP2007, savPPNP2013, braEPJA2015}, although 
this picture, {and the underlying collective nature of the mode, are} still under debate 
~\cite{reiPRC2013}. 

One major reason for the recently increased interest in the PDR is the possibility of carrying out 
several measurements on these low-lying dipole excitations, using heavy-ion~\cite{pelPLB2014, crePRC2015}, proton~\cite{polPRC2012, kruPLB2015}, and $\alpha$ inelastic scattering experiments~\cite{savPRL2006, endPRC2009}. Indeed, the experimental study of the PDR with different probes provides intimate information about the isospin nature of these excitations which would not be possible to infer from $\gamma$ experiments alone~\cite{braEPJA2015}.  
These experimental discoveries were followed by intensive theoretical investigations, focusing on the isoscalar (IS)/isovector (IV) character of dipole excitations in isospin-asymmetric nuclear systems~\cite{Papak2014,zhePRC2016}. 


In this paper we aim at getting a deeper insight into the features of the dipole response in nuclei, with a special attention to the role of neutron/proton imbalance. 
By looking at the dynamical response of the system to different kinds of external perturbations, we explore
{spatial profile and IS/IV character} of the dipole excitation modes in neutron-rich nuclei and Sn isotopes. 

This study is tackled by investigating 
the small amplitude limit of the dynamical nuclear response to a dipole operator within the quantal Time Dependent Hartree-Fock (TDHF) method~\cite{Lac04,Sim10,Nak16}, its zero amplitude limit, known as the Random-Phase Approximation (RPA)~\cite{Rin80}, as well as within its semi-classical analog: the Vlasov equation~\cite{Brink1986,Burgio1988,Matera}. 
 

As a key point, the paper presents, for the first time, a detailed comparison between TDHF and RPA calculations, with the purpose of verifying numerically
the analitical equivalence of the two approaches in the small amplitude limit. In such a way, we aim at bringing out the possible emergence of spurious differences arising from technical details and assessing practical 
advantages or drawbacks of the two procedures.

Examining analogies and differences between semi-classical and quantal results, 
one expects to learn more about the nature and the degree of collectivity of excitation modes of present experimental interest. 
A schematic interpretation of nuclear excitations in terms of collective motion, whenever possible, may allow one to establish a more direct connection to global features of the nuclear effective interaction, such as surface tension and symmetry energy,  also linking the nuclear response to macroscopic properties of nuclei, like density profile and neutron skin \cite{bar_pygmy2013}. 
To this purpose, we will also examine the sensitivity of the dipole response to specific ingredients of the nuclear mean-field potential and to the Equation of State (EoS), adopting Skyrme parametrizations which mainly differ in the isovector channel~\cite{SAMi-J}, already employed in recent structure studies 
{\cite{SAMi-ref}}. 

Hence, from our analysis, we also aim at extracting important information on some aspects of effective interaction and nuclear EoS of considerable relevance also in other fields, such as heavy-ion reactions
and nuclear astrophysics. Lastly, we note that, whereas we concentrate on 
dipole excitations in the present work, the same investigation can be extended to other multipolarities as well {(cf., e.g., Ref. \cite{Yuksel})}.

The paper is organized as follows: in Section II we introduce the approaches employed in our analysis and the details of the calculations related to the dipole response. Section III is devoted to the discussion of the results obtained  for dipole strengths and transition densities of selected neutron-rich nuclei and Sn isotopes. We discuss in particular the features of the low-lying region of the dipole response, showing interesting connections between isoscalar and isovector strengths in neutron-rich nuclei. The paper ends up with a summary and some perspectives in Section IV.   

\section{Theoretical framework}

\subsection{Microscopic approaches and effective interactions}
In the present work, we compare three different microscopic theories that are widely used to describe many-body dynamics: the TDHF, the RPA and the Vlasov approaches. 
Main features and connections among them are briefly discussed below.

In the TDHF theory, the evolution of the one-body density matrix ${\rho}(t)$ is determined by 
\begin{equation}
i\hbar \partial_{t}{\rho}(t)= \left [ h \left [{\rho} \right],{\rho}(t) \right ], 
\label{EQ:TDHF}
\end{equation}
where $h \left [{\rho} \right ]={\bf p}^{2}/2m+U \left [ {\rho}\right ]$ is the non-relativistic single-particle Hamiltonian with $U \left [{\rho} \right ]$ being the self-consistent mean-field potential, that is a functional of the 
one-body density.  

The TDHF approach is nowadays widely used in nuclear physics to describe various aspects of nuclear dynamics~\cite{Lac04,Sim10,Nak16,Sim18}. 
Here we will restrict ourselves to the study of small deviations from the equilibrium density $\rho_{e}$. 
The small amplitude fluctuations $\delta \rho (t) = \rho(t) - \rho_e$ can be determined either by solving explicitly the time-dependent evolution given by Eq.~(\ref{EQ:TDHF}) or by linearizing the TDHF equation, leading to the RPA approach. Keeping only terms linear in $\delta \rho$, Eq.~(\ref{EQ:TDHF}) equals to 
\begin{equation}
i\hbar \frac{\partial }{\partial t}\delta \rho = \left [ h \left [{\rho_e}\right ],\delta \rho \right ]+\left [\frac{\partial U}{\partial \rho}\cdot \delta \rho , \rho_e \right ],
\label{EQ:RPA-2}
\end{equation}
so 
one can access the response function of the system to a small external 
perturbation.
Therefore, despite the TDHF has a larger domain of applicability, from the analytical point of view, it is  
equivalent in the small amplitude regime to the RPA approach, where the time-evolution is replaced by an eigenvalue problem.  
A detailed discussion on the RPA method employed to obtain the results 
presented here can be found in~\cite{rpa2013} and references therein. 

The Vlasov equation, which describes the time evolution of the one-body distribution function in phase space, represents instead the semi-classical limit of TDHF and, for small-amplitude dynamics, of the RPA equations~\cite{Rin80}.
This self-consistent approach is suitable to describe robust quantum modes, of zero-sound type, in both nuclear matter and finite nuclei~\cite{barPR2005,Burgio1988,urbPRC2012,barPRC2012}, though it is unable to account for effects associated with the shell structure. 
Expliciting the two species constituting nuclear matter,
one has essentially to solve the two coupled Vlasov kinetic equations for the neutron and proton distribution functions $f_q({\bf r},{\bf p},t)$, with $q=n,p$~\cite{barPR2005}:
\begin{equation}
\frac{\partial f_q}{\partial t}+\frac{\partial \epsilon_q}{\partial {\bf p}}\frac{\partial f_q}{\partial {\bf r}}- \frac{\partial \epsilon_q}{\partial {\bf r}}\frac{\partial f_q}{\partial {\bf p}}=0. 
\label{vlasov}
\end{equation}
In the equations above, $\epsilon_q$ represents the neutron or proton single particle energy, which contains the mean-field potential $U_q$. 

To represent the nuclear effective interaction, we start considering a given energy density functional $\mathscr{E}[\rho]$. 
{
This formulation is very convenient since it allows to extract the mean-field potential as its functional derivative with respect to the density. Actually, the residual interaction, i.e., the anti-symmetrized particle-hole interaction used in RPA calculations,  is calculated via the second functional derivative of $\mathscr{E}[\rho]$, when dealing with density dependent forces.}


Considering a standard Skyrme interaction, 
{and specializing to even-even systems,}
the  functional $\mathscr{E}[\rho]$ is expressed in terms of the isoscalar, $\rho=\rho_n+\rho_p$, and isovector, $\rho_{3}=\rho_n-\rho_p$,  densities and 
kinetic energy densities ($\tau=\tau_{n}+\tau_{p}, \tau_{3}=\tau_{n}-\tau_{p}$) as~\cite{radutaEJPA2014}:
\begin{eqnarray}
\mathscr{E}&=&\frac{\hbar^2}{2 m}\tau + C_0\rho^2 + D_0\rho_{3}^2 + C_3\rho^{\alpha + 2} + D_3\rho^{\alpha}\rho_{3}^2 ~+ C_{eff}\rho\tau \nonumber\\
&& + D_{eff}\rho_{3}\tau_{3} + C_{surf}(\bigtriangledown\rho)^2 + D_{surf}(\bigtriangledown\rho_3)^2,
\label{eq:rhoE}
\end{eqnarray}
where $m$ is the nucleon mass and the coefficients $C_{..}$, $D_{..}$ are combinations of {the} Skyrme parameters~\cite{zheAXV2018}. 
{The actual Skyrme functional is more complicated than in Eq.
(\ref{eq:rhoE}) as it includes the spin-orbit terms, plus other terms
that depend on the spin-orbit densities $\vec J$ and are dubbed
$J^2$ terms.}
The spin-orbit terms are considered in TDHF and RPA calculations, but they are not included in the semi-classical Vlasov calculations.
{The $J^2$ terms are not included in the TDHF and Vlasov calculations. 
Although they could be included in RPA, 
for the sake of comparing it with other models, 
they are dropped in RPA as well.
One may expect that the overall qualitative features of the excitations 
investigated here are not significantly affected by the approximations 
we have made.}
The Coulomb interaction is considered in all frameworks. 
  
We are interested in effects linked to the neutron/proton content of the nuclei under study, thus it is convenient to introduce the symmetry energy per nucleon, $E_{sym}/ A = C(\rho) I^2$, where $I = \rho_3/\rho$ is the asymmetry parameter and the coefficient $C(\rho)$ can be written as a function of the Skyrme coefficients:
\begin{equation}
C(\rho) = \frac{\epsilon_F}{3} + D_0\rho + D_3\rho^{\alpha+1} ~+ 
\frac{2m}{\hbar^2}\left(\frac{C_{eff}}{3} + D_{eff}\right)\epsilon_F\rho,
\end{equation}
with $\epsilon_F$ denoting the Fermi energy at density $\rho$. 


In the following, we will adopt the recently introduced SAMi-J Skyrme effective interactions~\cite{SAMi-J} based on the fitting protocol of the SAMi interaction~\cite{SAMi}. 
The SAMi-J family has been produced by systematically varying the value of 
J = $C(\rho_0)$ (being $\rho_0$ the saturation density) from 27 to 35 MeV, keeping fixed the optimal value of the incompressibility and effective mass predicted by SAMi and refitting again the parameters for each value of J. This produces a set of interactions of similar quality on the isoscalar channel and that, approximately,  isolate the effects of modifying the isovector channel in the study of a given observable.
The SAMi fitting protocol~\cite{SAMi} includes: binding energies and charge radii of some doubly magic nuclei, which {allows} the SAMi-J family to predict a reasonable saturation density 
and energy  of symmetric  nuclear  matter 
(the incompressibility value is $K = 245$ MeV)  
; some selected spin-orbit splittings; the spin and spin-isospin Landau Migdal parameters~\cite{LMparam}; 
and, finally, the neutron matter EoS of Ref.~\cite{wir19}. 
These  features allow the new SAMi-J interactions to give a reasonable description of isospin as well as  spin-isospin  resonances, keeping a good reproduction of well known empirical data such as masses, radii and important nuclear excitations (see original work for further details). 

In our calculations, we employed three SAMi-J parametrizations: SAMi-J27, SAMi-J31 and SAMi-J35~\cite{SAMi-J}. Since, as mentioned above, the SAMi-J interactions have been fitted in order to also reproduce the main features of finite nuclei, for the three parametrizations the symmetry energy coefficient gets the same value, $C(\rho_c) \approx 22$ MeV at $\rho_c = 0.6\rho_0$, which 
would approximately correspond to the average density probed by nuclear masses via the fitting protocol, if one assumes a local density approximation \cite{roca-maza2015,roca-maza2018}.
The corresponding values of symmetry energy at saturation, together with the values of the slope parameter $\displaystyle L = 3 \left. \rho_0 \frac{d C(\rho)}{d \rho} \right\vert_{\rho=\rho_0}$ are reported in Table~\ref{sl}.
\begin{table}[h]
{\renewcommand\arraystretch{1.2}
\begin{tabular}{|c|c|c|}
\hline
Interaction  & J [MeV] &  L [MeV] \\
\hline
SAMi-J27 & 27 & 29.9  \\
\hline
SAMi-J31 & 31 & 74.5 \\
\hline
SAMi-J35 & 35 & 115.2 \\
\hline
\end{tabular}}
\caption{The symmetry energy coefficient at saturation density for the Skyrme interactions employed in our study and the corresponding slope $L$.}
\label{sl}
\end{table}

\subsection{Numerical details of the calculations and ground state configuration}
In order to determine the ground state configuration of the nuclei under study, different numerical procedures are followed in quantal and semi-classical approaches. In the quantal case, Hartree-Fock (HF) calculations are performed, although two different codes are employed for TDHF and RPA calculations. In the former case, the {\sc ev8} code~\cite{Bon2005} is used 
while in the latter case the code {\sc skyrme\_rpa}~\cite{rpa2013} is employed. In the present study, we consider selected closed-shell nuclei and some Sn isotopes known to be spherical (see e.g.~\cite{moller2016}). 
{In addition, pairing correlations 
have been neglected in order to allow for a consistent comparison between the different approaches. 
{Pairing will not play a role in the magic nuclei that we discuss
below. In the open-shell spherical systems, pairing is known to affect
more the low-lying quadrupole and octupole states than the dipole response
~\cite{ScaPHD, Sca13}.
}
{More specifically, in Ref.\cite{polar} it has been shown that pairing effects have no influence on the dipole polarizability in the ${}^{116-132}$Sn isotopes, especially in the case of SAMi-J31, that is employed here}. 

In the {\sc ev8} code, the HF equations are solved in coordinate space. The mesh size has been taken as $dx = 0.8$ fm. The imaginary time method is adopted, 
with a fixed time step $dt_0 = 0.36$ fm/c.
These parameters correspond to standard choices~\cite{Bon2005}. 
The total size of the cubic mesh {should be large enough to avoid effects of particle evaporation on the TDHF dynamical response}. We will consider several choices to test the sensitivity of the 
results to this parameter.  
However, the values considered should also ensure a reasonable computational time. 

On the other side, within the fully self-consistent HF+RPA calculations \cite{rpa2013} presented here, the ground state properties of the different nuclei are calculated in coordinate space using box boundary conditions {assuming spherical symmetry}. 
Also in this case, we will test different sizes of the box, 
{keeping a radial mesh of dr = 0.1 fm}. The same box is used to calculate discrete states at positive energy that are associated with the continuum part of the spectrum. A cutoff energy of 120 MeV (in the single-particle energy) is adopted for the RPA calculations. With this energy cutoff, we have checked that the energy weighted sum rule is satisfactorily fulfilled. 


The integration of the Vlasov transport equations is based on the test-particle (t.p.) (or pseudo-particle) method~\cite{wong}, with a number of $1500$ t.p. per nucleon in all the cases, ensuring in this way a good spanning of the phase space. The ground state configuration corresponds to the stationary solution of Eq.~(\ref{vlasov}). Within the Thomas-Fermi (TF) approximation, we adopt the following numerical procedure: neutrons and protons are distributed inside spheres of radii $R_n$ and $R_p$, respectively. Accordingly, particle momenta are initialized inside Fermi spheres associated with the local neutron or proton densities.  Then   $R_n$ and $R_p$ are tuned in order to minimize the corresponding total energy, associated with the effective interaction adopted in the calculations. Because test particles are often associated with finite width wave packets (we use triangular functions~\cite{TWINGO}), some surface effects are automatically included in the initialization procedure and in the dynamics, even though explicit surface terms, as those contained in the effective Skyrme interactions, are not considered. This implies that, for the surface terms, one cannot simply use the coefficients associated with the SAMi-J parametrizations. Indeed we observe that a good reproduction of the experimental values of both proton root mean square radius and binding energy, for the nuclei selected in our analysis, is obtained when taking  $C_{surf} = D_{surf} = 0$ in our parametrizations. 
Thus this choice has been adopted in the following for Vlasov calculations (see Ref.~\cite{zhePRC2016} for more details).    


\subsection{Dipole response}
\label{sec:dipole}
Dipole oscillations and response functions can be investigated, in both TDHF and semi-classical treatments, introducing a small perturbation of the ground state configuration of the nucleus under study and then looking at its dynamical evolution, as given by Eq.~(\ref{EQ:TDHF}) or Eq.~(\ref{vlasov}). Thus we study the E1 (isoscalar and isovector) response of nuclear systems, considering initial conditions determined by the instantaneous excitation $\displaystyle V_{ext} =\eta_k \delta(t-t_0) \hat{D}_k$, 
along the $z$ direction~\cite{calAP1997,barPRC2012}. 
Here $\hat{D}_k$ denotes the operator employed to introduce isoscalar ($k=$ S) or isovector ($k=$ V) dipole excitations {and takes the standard form~\cite{rpa2013}}: 
\begin{align}
\hat{D}_S &= \sum_{i=1}^A \left (r_i^2 - \frac{5}{3} \langle r^2\rangle \right )z_i, \label{dis} \\
\hat{D}_V &= \sum_{i=1}^A \left [ \tau_i \frac{N}{A} - \left (1-\tau_i \right ) \frac{Z}{A} \right ] z_i, \label{div}
\end{align}
where $N$ and $Z$ indicate neutron and proton number, $A = N+Z$, $\tau_i =1(0)$ for protons (neutrons) and  $\langle r^2\rangle$ denotes the mean square radius of the nucleus considered. The above definitions [Eqs.~(\ref{dis}) and (\ref{div})] avoid the undesired effect of the so called spurious state and remove the contribution from the center of mass, respectively. We note that the operator $\hat{D}_V$ also contains an isoscalar component, which vanishes only for symmetric (N = Z) systems. According to basic quantum mechanics, if $|\Phi_{0} \rangle$ is the state before perturbation, then the excited state becomes $\displaystyle |\Phi_k (t_0)\rangle =e^{i \eta_k \hat{D}_k} |\Phi_{0} \rangle$. The value of $\eta_k$ can be related to the initial expectation value of the collective dipole momentum $\hat{\Pi}_k$, which is canonically conjugated to the collective coordinate $\hat{D}_k$, i.e.,  $[\hat{D}_k,\hat{\Pi}_k]=i\hbar$~\cite{barRJP2012}.


The same operators defined above are considered in RPA calculations, to extract isoscalar and isovector dipole strength functions: $S_k(E)=\sum_{n > 0}|\langle n|\hat{D}_k|0\rangle|^2\delta(E-(E_n-E_0))$, 
where $E_n$ is the excitation energy of the state $|n\rangle$ and $E_0$ is the energy of the ground state $\displaystyle |0\rangle=|\Phi_{0} \rangle$.

In TDHF and Vlasov calculations, the strength function is obtained from the imaginary part of the Fourier transform of the time-dependent expectation value of the dipole moment $ \displaystyle D_k(t) = \langle \Phi_k (t) |\hat{D}_k| \Phi_k (t) \rangle $ as:
\begin{equation}
S_k(E) =\frac{Im (D_k(\omega))}{\pi \eta_k },
\label{stre}
\end{equation}
where $\displaystyle D_k(\omega) =\int_{t_0}^{t_{max}} D_k(t) e^{i\omega t} dt$, with $E = \hbar\omega$.  
In these two approaches, we follow the dynamics of the system, looking 
in particular at the time oscillations of the dipole moments,  
until $t_{max}=1800$~fm/c.
The TDHF equations are solved using the 3D-TDHF code of Refs.~\cite{Kim97,Lac02,Sca13,Sca14}, with  a time step $dt = 0.36$ fm/c.
A slightly larger time step, {$dt = 0.50$ fm/c}, is instead adopted for the solution of the Vlasov equation.  
As described in~\cite{reiPRE2006}, in order to eliminate the artifacts resulting from a finite time domain analysis of the signal, a filtering procedure was moreover applied by introducing a smooth cut-off function such that
\begin{equation}
D_k(t) \rightarrow D_k(t) \exp \left (-\frac{\gamma t}{2 \hbar} \right ),
\label{eq:smooth}
\end{equation}
with $\gamma = 0.8$ MeV.

\section{Results}
This section is dedicated to investigate
the E1 (IS and IV) response of neutron-rich nuclear systems. In order to compare with the semi-classical results reported in a previous work~\cite{zhePRC2016}, we consider in our analysis three closed-shell nuclei: $^{68}$Ni (proton closed-shell), $^{132}$Sn and $^{208}$Pb. 
{Later,} to better explore how the features of the dipole response evolve with the neutron/proton content of the nuclei
under study, we will also  consider two other Sn isotopes ($^{100}$Sn, $^{120}$Sn) in our analysis. 

\subsection{Comparison between quantal and semi-classical approaches}
\label{sec:comparison_classical_quantal}
\paragraph*{\textbf{\textup{\noindent Ground state properties and density profiles.}}}
As stressed in the Introduction, we aim at elucidating the role of some global properties, such as density profiles and neutron skin, in determining the main features of the nuclear response. Therefore, as a preliminary step, it is worthwhile to illustrate the capability of both semi-classical and quantal approaches in reproducing some experimental ground state quantities. {It should be noticed that the HF calculations give an excellent agreement with data if the $J^2$ terms are included. Indeed the SAMi family has been originally fitted including all Skyrme-like terms.} 

\begin{figure}[b]
\includegraphics[scale=0.3]{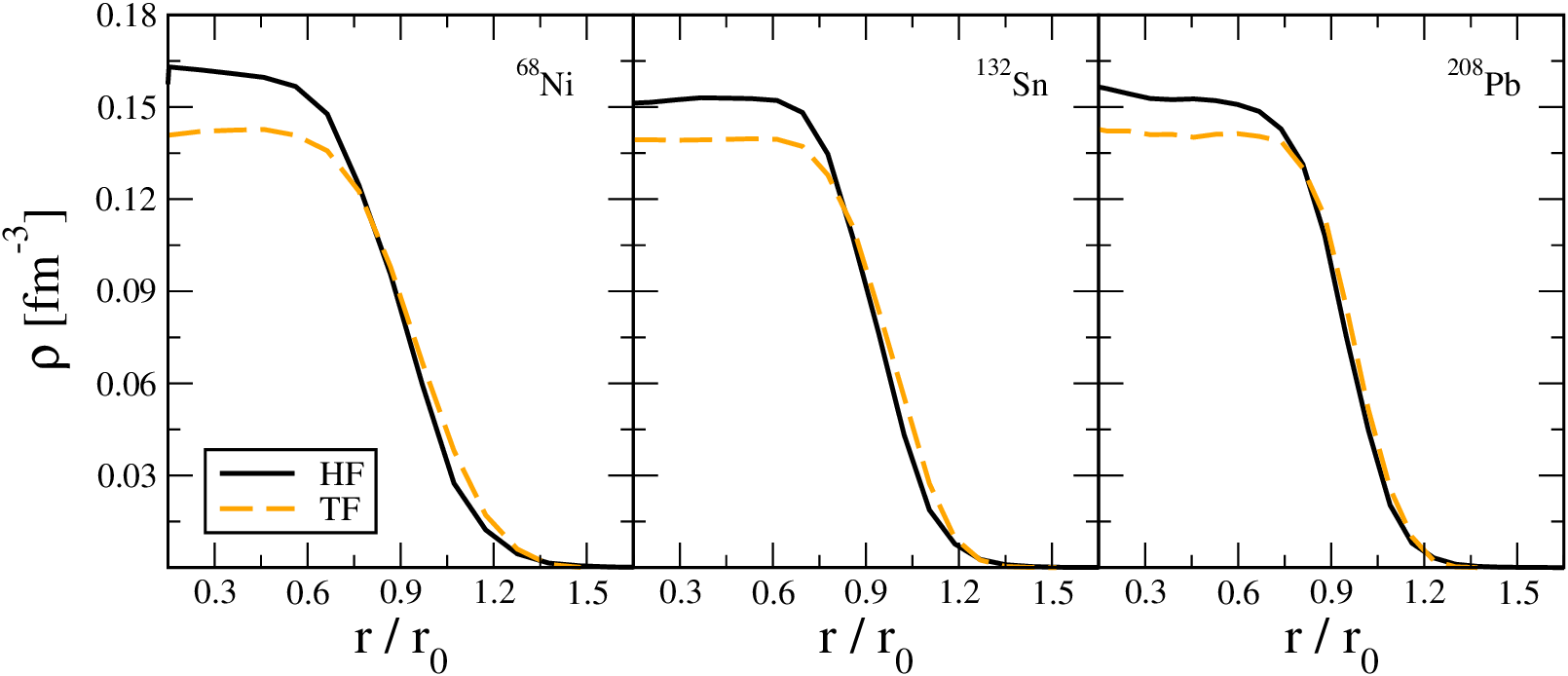}
\caption{(Color online) The isoscalar density profiles 
for $^{68}$Ni, $^{132}$Sn and $^{208}$Pb, from Hartree-Fock and Thomas-Fermi models, versus the normalized radius $r/r_0$, with $r_0 = 1.2~A^{1/3}$.}
\label{fig03} 
\end{figure}
The proton root mean square radius and the binding energy evaluated by employing, respectively, a semi-classical treatment in the TF approximation 
or a self-consistent quantal HF calculation, 
are listed in Table~\ref{tab:skin}, together with the corresponding experimental values. For the sake of completeness, neutron root mean square radius and neutron skin thickness are also reported. The SAMi-J31 parameterization of the effective interaction has been employed. 

\begin{table}[t]
{\renewcommand\arraystretch{1.2}
\begin{tabular}{c|cccc}
\hline
\hline 
& \textbf{$\sqrt{\langle r^2 \rangle_n}$} [fm] &  \textbf{$\sqrt{\langle r^2 \rangle_p}$} [fm] &  \textbf{$\sqrt{\langle r^2 \rangle_n} -\sqrt{\langle r^2 \rangle_p}$} [fm] & \textbf{$\frac{B}{A}$} [MeV] \\
\hline
\hline
\multicolumn{5}{c}{\textbf{$^{68}$Ni}} \\
\hline
HF & 4.001 & 3.831 & 0.170 & 8.845 \\
\hline
TF & 4.102 & 3.898 & 0.204 & 9.050 \\
\hline
Exp & ---  & 3.857  & --- & 8.682  \\
\hline
\hline
\multicolumn{5}{c}{\textbf{$^{132}$Sn}} \\
\hline
HF & 4.927 & 4.664 & 0.263 & 8.448  \\
\hline
TF & 5.035 & 4.741 & 0.294 & 8.552  \\
\hline
 Exp & ---  & 4.709 & --- & 8.354  \\
\hline
\hline
\multicolumn{5}{c}{\textbf{$^{208}$Pb}}  \\
\hline
HF & 5.654 & 5.456 & 0.198 & 7.916 \\
\hline
TF & 5.735 & 5.536 & 0.199 & 8.042 \\
\hline
Exp & ---  & 5.501 & ---  & 7.867  \\
\hline
\hline
\end{tabular}}
\caption
{Neutron and proton root mean square radii, and their difference, and binding energy for three systems considered in our study, as obtained in TF and HF models with the SAMi-J31 interaction. The experimental values, for charge radius and binding energy, are also indicated \cite{dataref}.
} 
\label{tab:skin}
\end{table}

One observes, in both models, a general good reproduction of the experimental values, especially for larger systems, as it should be, according to the mean-field approximation adopted. 
TF calculations predict a more extended neutron skin as well as slightly larger binding energy values with respect to the HF case. 
To better emphasize the differences observed between the two approaches, the isoscalar density $\rho$ and the local asymmetry $\rho_3/\rho$ profiles 
are plotted in Figs.~\ref{fig03} and \ref{fig02}, respectively. 

\begin{figure}[b]
\includegraphics[scale=0.3]{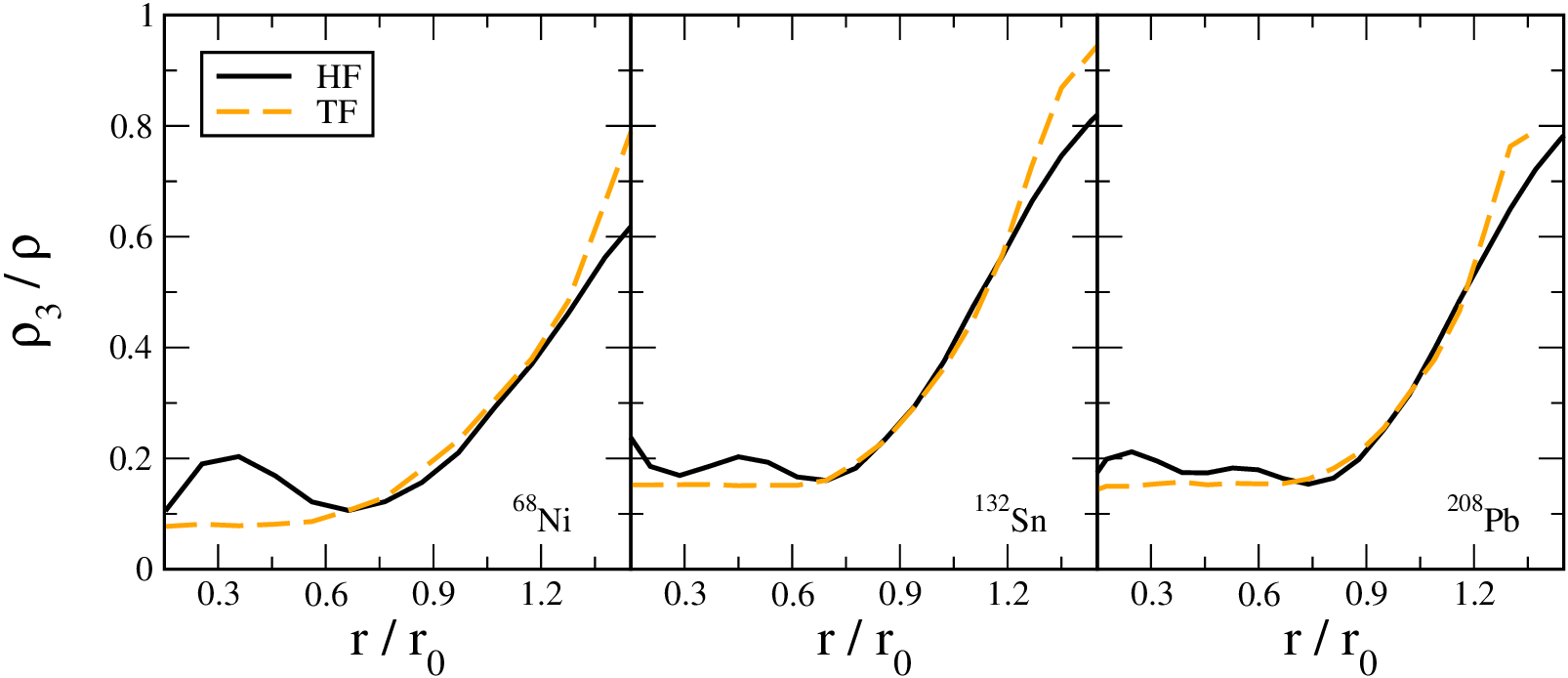}
\caption{(Color online) The {local} asymmetry profile 
for $^{68}$Ni, $^{132}$Sn and $^{208}$Pb, from Hartree-Fock and Thomas-Fermi models,  versus the normalized radius $r/r_0$, with $r_0 = 1.2~A^{1/3}$.}
\label{fig02} 
\end{figure}
With respect to the HF result, the TF isoscalar density profile appears flatter in the internal region, especially in the Ni and Pb case, indicating a sharper transition from the volume to the surface region.  
This could be attributed to the numerical treatment of surface effects in Vlasov calculations and to the lack of intrinsic quantal gradient terms{, corresponding to the $\hbar^2$ terms in the Wigner-Kirkwood $\hbar$-expansion of the distribution function~\cite{JenPLB78, SouNPA2000}}. One expects that these differences will affect the details of the modes mostly 
involving surface oscillations.
Looking at Fig.~\ref{fig02}, one observes some differences  between quantal and semi-classical predictions also in the isovector density $\rho_3$. 
{Semi-classical calculations are characterized by a larger neutron drift towards 
the surface. Some differences appear also in the more internal region, evidencing the role of shell effects in shaping the fine details of the nuclear structure. }


\paragraph*{\textbf{\textup{Dipole response and strength function.}}}

Next, we investigate the dipole response.
\begin{figure}[b]
\includegraphics[scale=0.3]{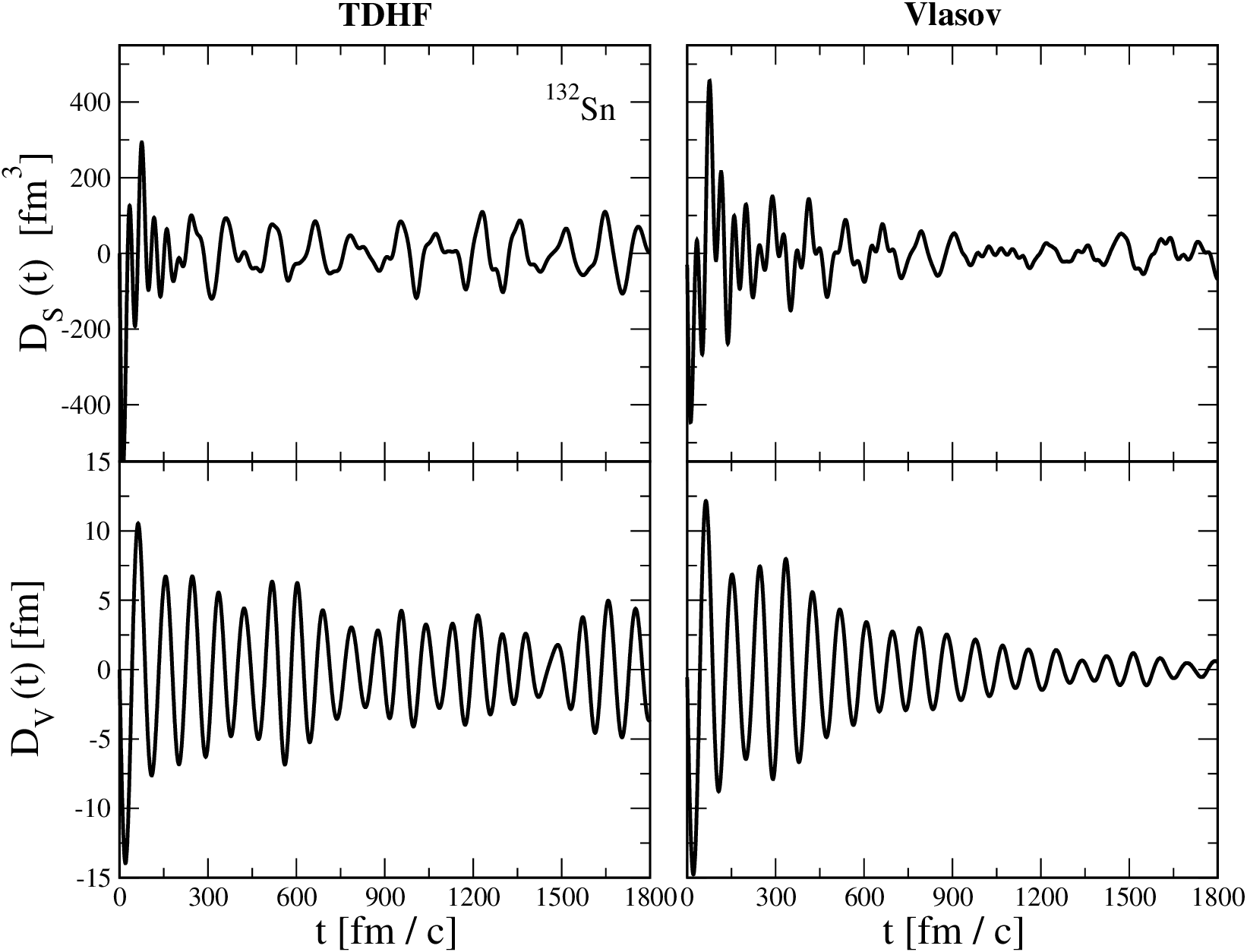}
\caption{(Top panels) The time evolution of the IS dipole moment for $^{132}$Sn and SAMi-J31 interaction, as obtained in TDHF (left panel) and Vlasov (right panel) calculations. 
(Bottom panels) The same as the top panels, but for the IV dipole moment. {TDHF dipole moments have been rescaled to the Vlasov perturbation strength.} }
\label{fig01_1} 
\end{figure}
Fig.~\ref{fig01_1} shows the time evolution of IS and IV dipole moments in the system $^{132}$Sn, as obtained by using an initial IS or IV { 
perturbation}. In our analysis we choose, as perturbation strength, the following values: $\eta_S = \e{1.0e-4}$ fm$^{-3}$, $\eta_V = \e{1.0e-4}$ fm$^{-1}$ in TDHF calculations 
and $\eta_S = \e{2.5e-3}$ fm$^{-3}$, {$\eta_V = \e{1.3e-1}$} fm$^{-1}$ for Vlasov ones, respectively.  The numerical procedure adopted to solve the 
Vlasov equation, related to the use of a finite number of test particles to map the one-body distribution function, introduces some numerical noise, implying to consider larger amplitude perturbations, with respect to TDHF.
One may generally note larger damping effects in the Vlasov calculations, probably related to the finite number of test particles and to the larger amplitude of the initial perturbation, that may induce non-linear effects, i.e. the coupling to other multipoles, and increase particle evaporation.   
  
\begin{figure*}
\includegraphics[width=0.3\textwidth,clip=true]{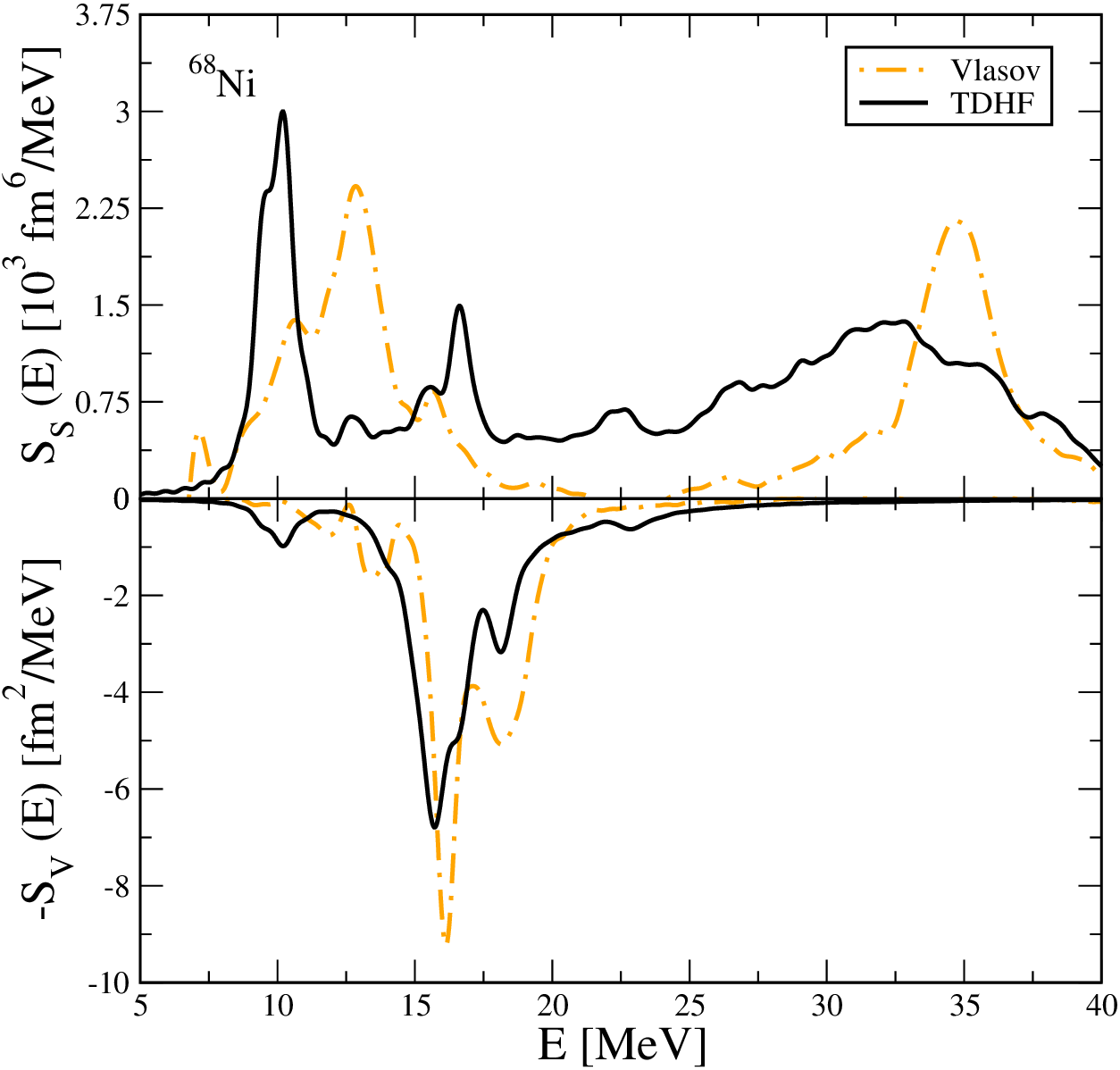} \quad \includegraphics[width=0.3\textwidth,clip=true]{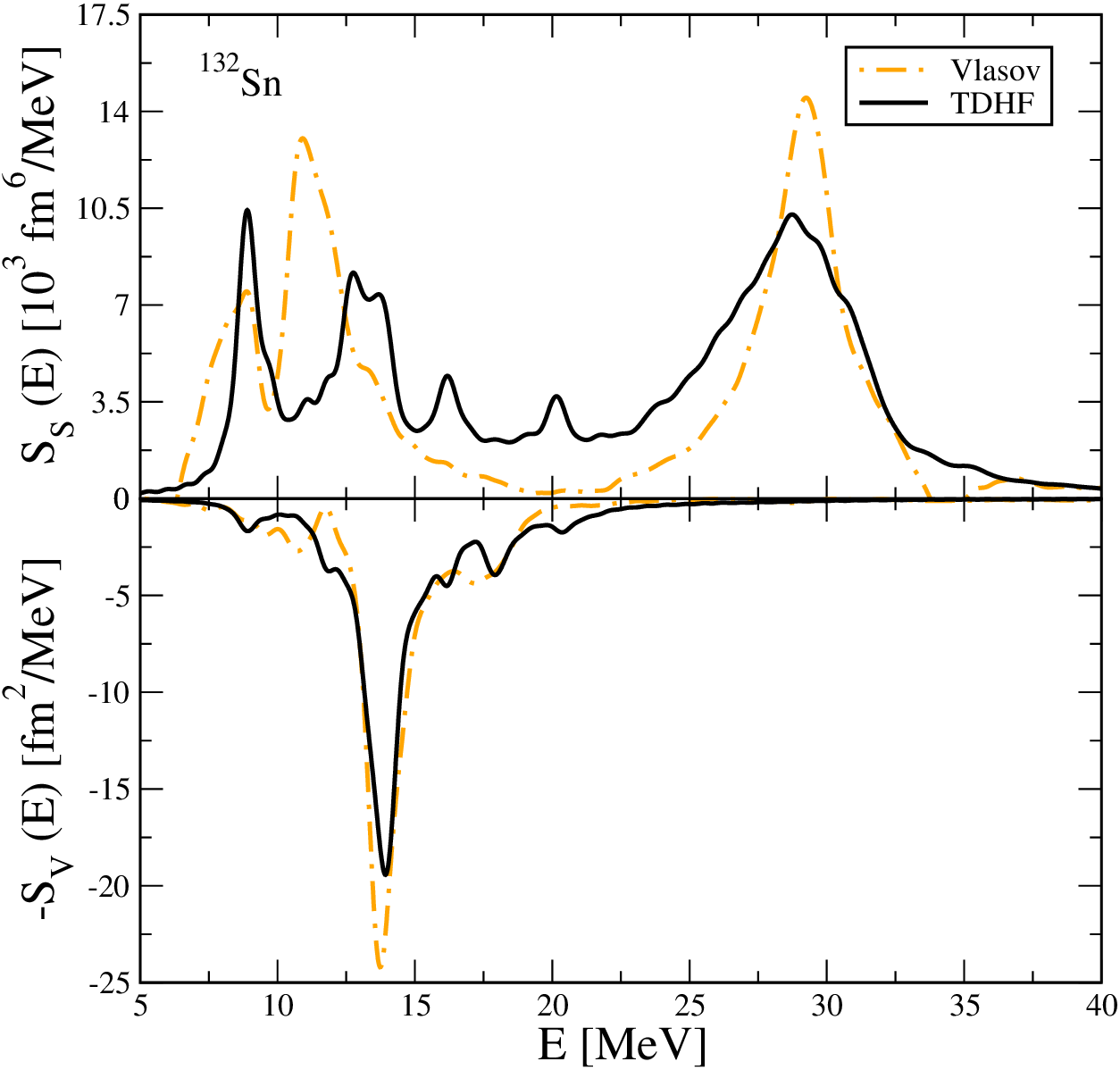} \quad \includegraphics[width=0.3\textwidth,clip=true]{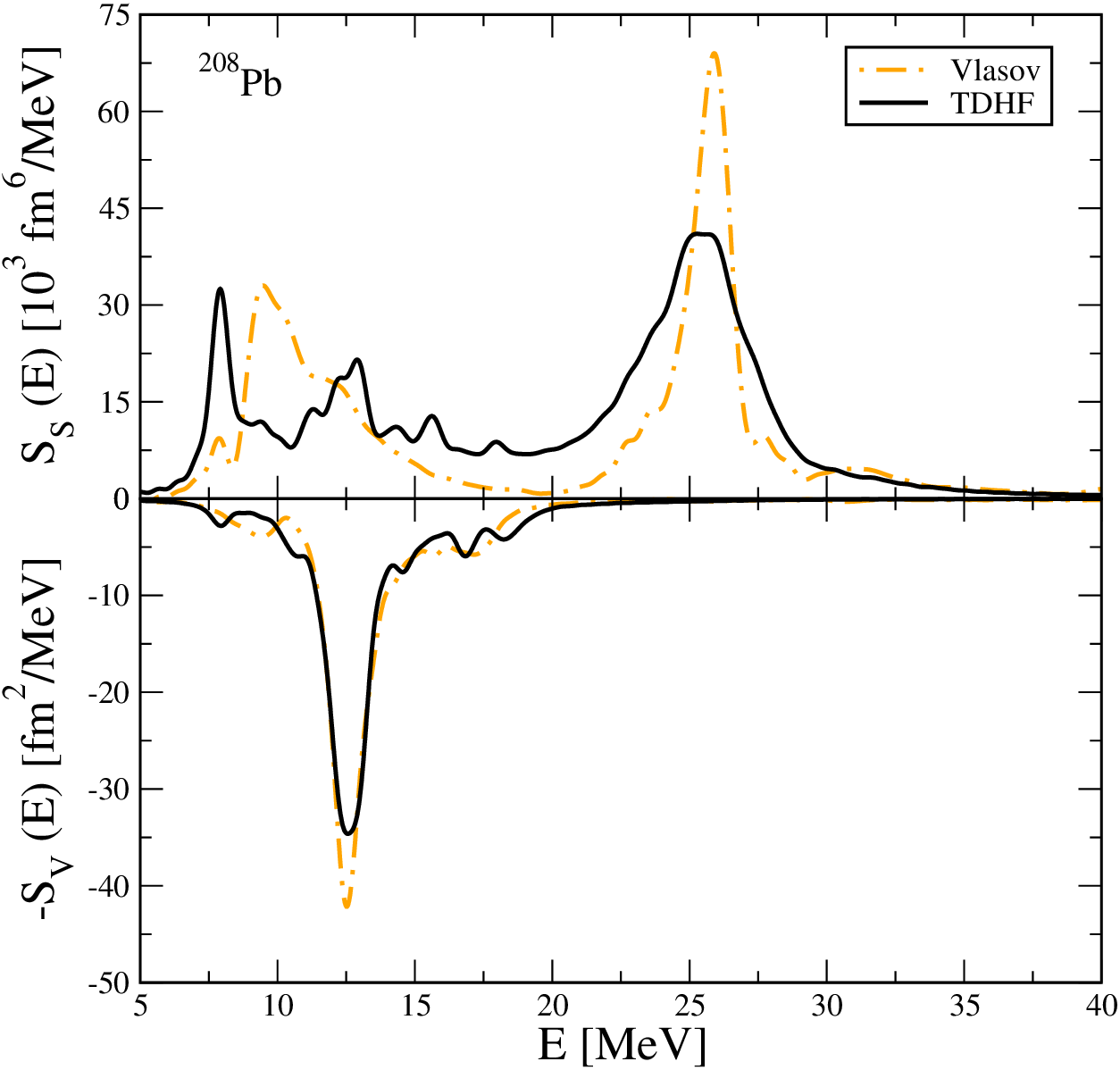}
\caption{(Color online) (Top panels) The strength function of the IS response of the closed shell nuclei $^{68}$Ni, $^{132}$Sn and $^{208}$Pb, with SAMi-J31 interaction. 
Full lines represent the results from TDHF calculations, dashed lines the ones from Vlasov calculations. (Bottom panels) Similar to the figures in the top panels, but for the strength function of the IV response.}
\label{fig01} 
\end{figure*}
Fig.~\ref{fig01} displays the corresponding IV and IS dipole strength functions in $^{68}$Ni, $^{132}$Sn and $^{208}$Pb. Looking at the bottom panels, one notices that the IV dipole strength is clearly dominated by the collective IV GDR mode peaked in the energy region around 12-16 MeV. Despite the differences 
observed between the semi-classical and quantal ground state features, the IV dipole strength deduced within the TDHF model is generally well reproduced by the corresponding Vlasov calculation. The agreement is particularly satisfying for the energy of the main IV peak, especially when larger systems are considered. Some strength is observed at low energy, 
i.e. in the region of the PDS 
{, albeit} the corresponding peaks look shifted in Vlasov calculations, with respect to TDHF. 
Though it is interesting to notice that the PDS also emerges in semi-classical calculations, we stress that the details of this low-energy IV contribution, namely its degree
of collectivity and precise energy location, are strongly affected by shell effects and by the ingredients of the residual interaction, as pointed out in recent investigations \cite{mazPRC2012,vret2012}.} 
 

Concerning the IS dipole strength (top panels in Fig.~\ref{fig01}), the scenario is more complicated. In the case of  the large systems, the two models give close values for the centroid of the high energy peak, in the excitation energy domain of the IS GDR (around 30 MeV), where the compressional modes dominate~\cite{Repko2013}. For $^{68}Ni$, a shift to higher energy is clearly evidenced in the Vlasov case, with respect to TDHF. Moreover, TDHF calculations lead to a larger fragmentation of the strength function, particularly in the Ni case. 
These differences may arise from the fact that in Vlasov  
simulations the evaporated particles are more abundant 
and may leave the calculation box. 
A remarkable discrepancy can be identified in the shape of the nuclear response in the low-energy regime. 
Two main regions of contribution can be recognized, which are well separated in energy in the TDHF case (see for instance the
contributions around 8 MeV and 13 MeV in the  $^{132}$Sn case). 
In the same energy region, the Vlasov calculations show two main peaks which are closer in energy (around 8 MeV and 11 MeV for  $^{132}$Sn) and not always clearly distinguishable (see in particular the result for $^{208}$Pb).  Moreover, the relative weight of the two peaks is
different in TDHF and Vlasov calculations.

It is worthwhile to notice that in previous {semi-classical} studies~\cite{urbPRC2012}, where isoscalar toroidal excitations were investigated, the modes emerging in this energy region have been associated with surface oscillations.  In particular, the lowest energy one corresponds to oscillations deeply 
involving the outer surface zone, whereas the second mode (of higher energy) would correspond to  standard toroidal dipole excitations, associated with the oscillation of the surface against the bulk region~\cite{urbPRC2012}. 

As anticipated above, the energy position {predicted for} these 
isoscalar surface peaks is quite different in Vlasov and TDHF calculations. 
In particular, in TDHF calculations, the second region of considerable strength is
shifted to higher frequency.  This discrepancy could originate
from  the {approximations done in the semi-classical approach, like
the lack of gradient terms and, on the other hand,} from the numerical procedure adopted to treat surface effects in this case, as already noted for the ground state configuration.   Since isoscalar gradient terms give a positive contribution to the restoring force, we may expect higher oscillation frequencies in TDHF. 
{Last but not least, the details of the low-lying excitations 
are very likely to be also affected by shell effects.}

Moreover, TDHF calculations seem to favor the lowest energy peak, whereas the opposite happens in the Vlasov results.
This latter behavior could be connected to the different density profile predicted by the two calculations.
A sharper evolution from the bulk to the surface region, as observed in the Vlasov case, seems to favor the dominance of the standard toroidal mode. On the other hand, a smoother density profile enhances surface effects, leading 
to more robust oscillations in the lowest-energy region. 
A further insight about the volume/surface nature of these excitation is gained, however, by looking at the shape of the corresponding transition densities, as discussed in the following (see Section~\ref{sec:comparison_tdhf_rpa}). 

\subsection{Isoscalar-isovector mixing in n-rich systems}
\begin{figure}[t]
\includegraphics[scale=0.3]{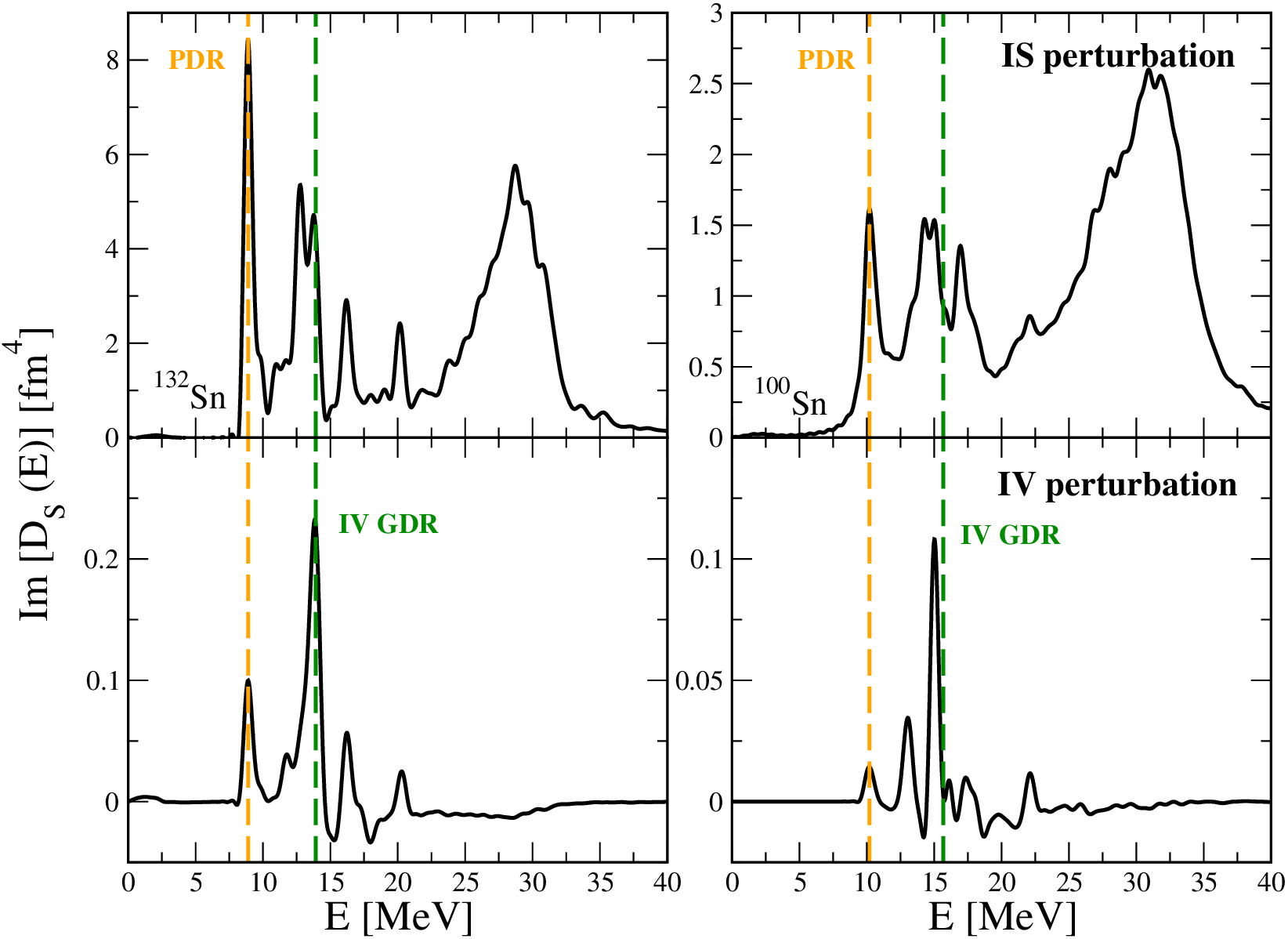}
\caption{(Color online) The imaginary part the of the IS response with IS perturbation (top) and IV perturbation (bottom), respectively, as obtained for $^{132}$Sn (left panels) and $^{100}$Sn (right panels) with SAMi-J31 in TDHF calculations.}
\label{fig06} 
\end{figure}

\begin{figure}[t]
\includegraphics[scale=0.3]{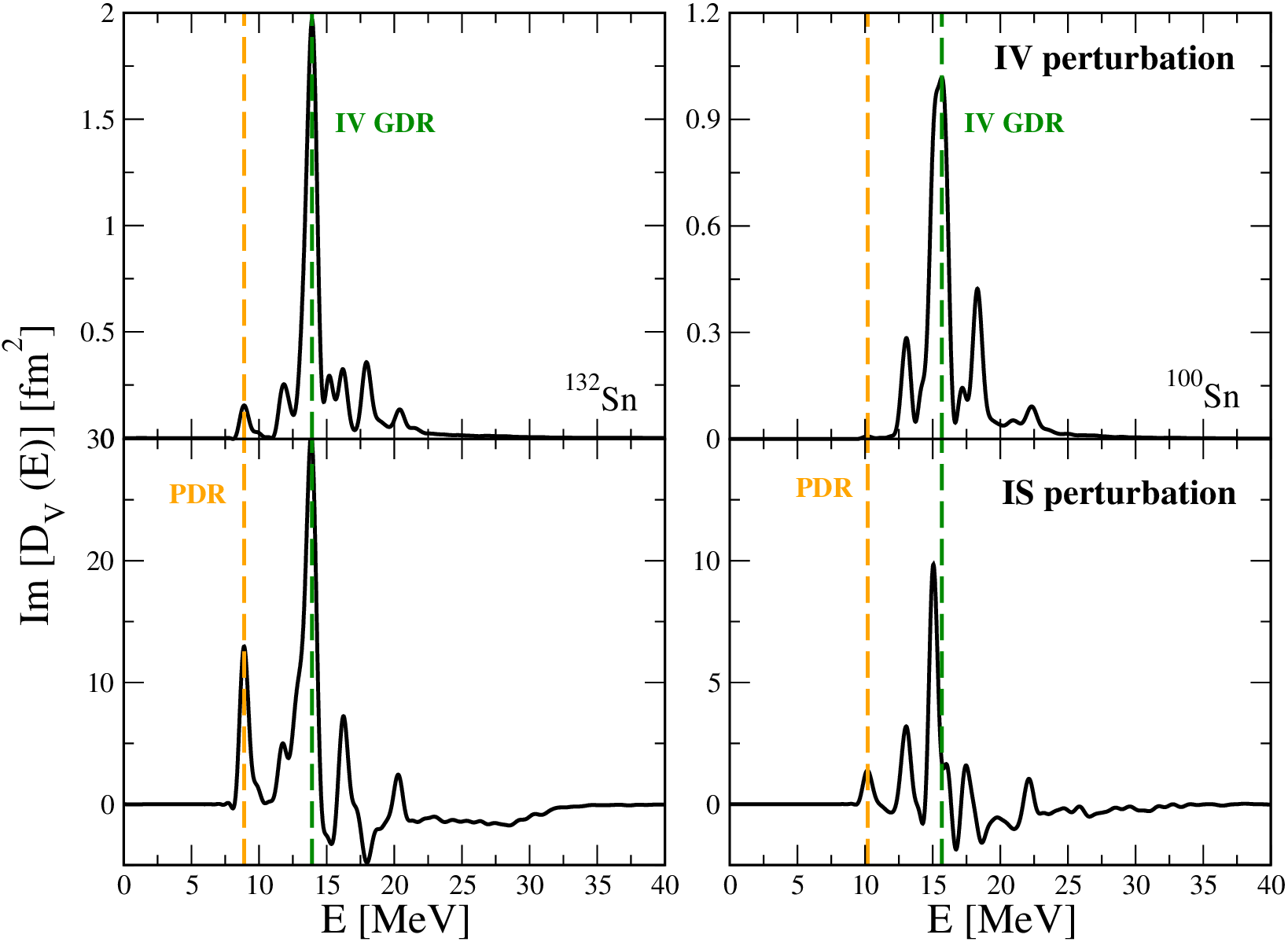}
\caption{(Color online) The imaginary part the of the IV response with IV perturbation (top) and IS perturbation (bottom), respectively, as obtained for $^{132}$Sn (left panels) and $^{100}$Sn (righ panels) with SAMi-J31 in TDHF calculations.}
\label{fig07} 
\end{figure}

As stated in the introduction, one of the goals of our analysis is to get a deeper insight into the isoscalar/isovector mixing which characterizes the excitation modes of nuclei with an unbalanced number of protons and neutrons. Let us consider 
TDHF calculations for the system $^{132}$Sn, with the SAMi-J31 effective interaction. 
Fig.~\ref{fig06} (left panels) shows that it is possible to extract a sizeable IS response by perturbing the nucleus not only with an initial 
IS excitation (top), but also by employing an initial IV excitation (bottom). 
Similarly, the investigation of the IV response carried out by employing the two kinds of initial perturbation (see Fig.~\ref{fig07}, left panels) shows that
IS excitations also generate an IV strength. 
{In each panel of Figs.~\ref{fig06}-\ref{fig07},} the relative height of the peaks will depend on the initial perturbation type
and on the intrinsic structure of the mode considered.

This characterization holds for almost all the main modes, which are excited by both the IS and IV perturbations, according to their mixed nature. 
Looking at the strength of the peaks in the different panels, it appears that, 
whereas the IV GDR manifests its well established predominant isovector nature, though with some mixing, the lowest-energy excitation (indicated as PDR) turns out to be mostly {an isoscalar-like} mode (see also the analysis in Ref.\cite{mazPRC2012}), which however can be excited also by an IV perturbation, owing to the coupling existing between isoscalar and isovector  modes in asymmetric systems. 
Only for the IS GDR, the IV projection comes out to be negligible, thus indicating a quite robust isoscalar nature of this mode. 

Looking at the bottom panels of Figs.~\ref{fig06} - \ref{fig07},  one can notice, as a quite interesting detail, that the two mixed projections{, i.e. 
the IS(IV) response generated by an IV(IS) perturbation} have the same structure. 
This confirms the consistency of our calculations.
It should be noticed that similar results about the isoscalar/isovector mixing
of the excitation modes in neutron-rich systems have been obtained also in 
semi-classical calculations \cite{zhePRC2016}. 

To better clarify the role of the isospin asymmetry in shaping the mixing observed in the dipole response, we extended our analysis to the nuclear system $^{100}$Sn, 
that is the double-magic nucleus of the Sn isotope chain, which is constituted by an equal number of protons and neutrons. 

According to the framework depicted above, one expects the coupling between the modes excited by the two perturbation kinds to be quenched in this case. 
Indeed, this is in line with the results plotted in {the right panels of Figs.~\ref{fig06} - \ref{fig07}}. 

Beyond dispute, in this case, the correspondence in the dipole strengths associated with the two perturbations considered is reduced. {For instance, the PDS is not observed, i.e. there is no IV strength in the PDR region
in the case of IV excitations (top right panel of Fig.~\ref{fig07}).}
 This stresses once again the prominant isoscalar nature of the pygmy mode and the strong connection of its IV counterpart with the neutron richness of the nucleus considered.
In the same way, owing to its isovector nature, the IV GDR peak (see the green line in the right panels of Figs.5-6) has a reduced strength in the IS response obtained with an IS perturbation  (top right panel of Fig.~\ref{fig06}). 
However, even for the symmetric $^{100}$Sn, the cross responses (right bottom panels) evidence the presence of some IS/IV mixing, mainly for the modes located in the  region between the PDR and the IV GDR. {
In particular,  
a noticeable degree of mixing is observed just slightly below the IV GDR,
reflecting a sudden transition from IS to IV excitations.
Therefore, even though the scenario for $^{100}$Sn is partially simplified 
by its N=Z nature, the general picture has not a trivial interpretation overall. The mixing observed 
{arises from} the Coulomb interaction, which breaks the symmetry between neutron and proton response.
}

For the following analyses  we will concentrate only on the IV (IS) response induced by an IV (IS) perturbation.

\subsection{Sensitivity to the effective interaction}
\begin{figure}[b]
\includegraphics[scale=0.3]{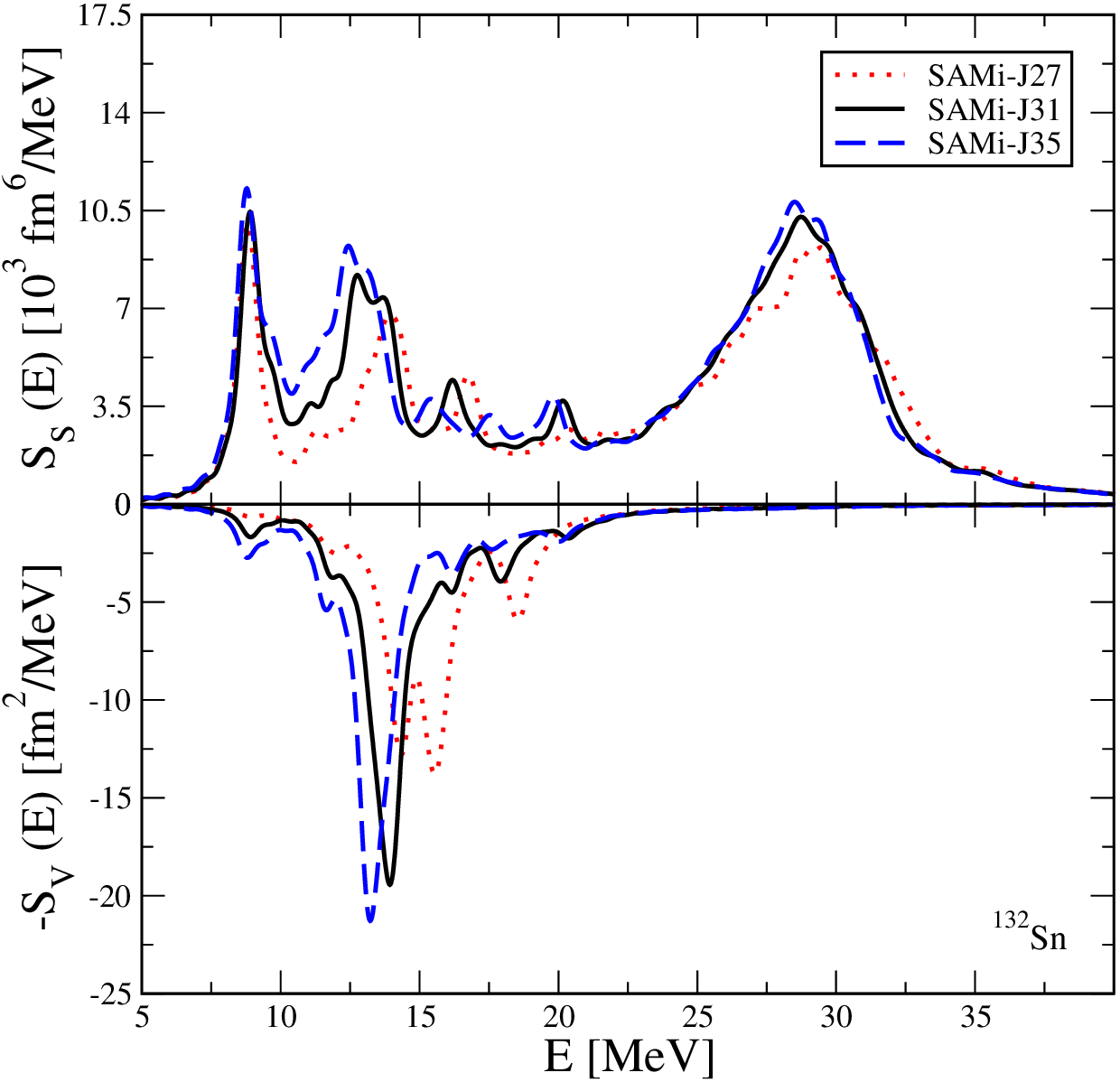}
\caption{(Color online) The IS and IV dipole response for $^{132}$Sn, as obtained 
in TDHF calculations when employing three SAMi-J parametrizations.}
\label{comparison_TDHF_samiall} 
\end{figure}

To discuss the impact of the employed effective interaction on the dipole response, we show in Fig.~\ref{comparison_TDHF_samiall} the results obtained for $^{132}$Sn, using three SAMi-J parametrizations differing by the (J-L) combination values (see Table~\ref{sl}). It is well known~\cite{trippa,carPRC2010} that the IV response is quite sensitive to the symmetry energy details, as we also observe here. In particular, the strength in the region below the IV GDR increases with L~\cite{piePRC2011, mazPRC2012}. We notice that, within the adopted interactions, also the neutron skin thickness increases with L (see Table~\ref{sl} and \ref{tab:skin}).
Also the frequency of the IV GDR is affected, and it moves to higher values as L decreases, reflecting the larger value of the symmetry energy at 
low density (below $\rho_c$). A splitting of the resonance in two peaks occurs
in the case of the SAMi-J27 interaction.  

On the other hand, we observe only a slight sensitivity of the IS response to the interactions considered in our study. The shift observed for the second relevant IS peak is probably related to the different isoscalar surface terms of the SAMi-J interactions considered, whose strength decreases from J27 to J35. 
{This observation supports the important impact of surface terms on the features of this mode, as already discussed when comparing TDHF and Vlasov results.} The compressional IS GDR is insensitive to the choice adopted  for the interaction, as one would expect considering the SAMi-J parametrizations are characterized by the same compressibility value.   




\subsection{Comparison between TDHF and RPA}
\label{sec:comparison_tdhf_rpa}

In this section, we aim at undertaking a detailed comparison of the dipole response which is extracted within the two quantal approaches employed in our work: TDHF and RPA. The two models are equivalent from the theoretical point of view, 
{at least in the limit of small oscillations}, so this analysis allows one to highlight possible spurious effects introduced by the technical procedure adopted and therefore isolate only the relevant physical features. Moreover, this study helps in giving some hints concerning the numerical parameters to be adopted to ensure the best possible agreement between the two codes, which could be used as a reference also for future works. 

\begin{figure}[t]
\includegraphics[scale=0.3]{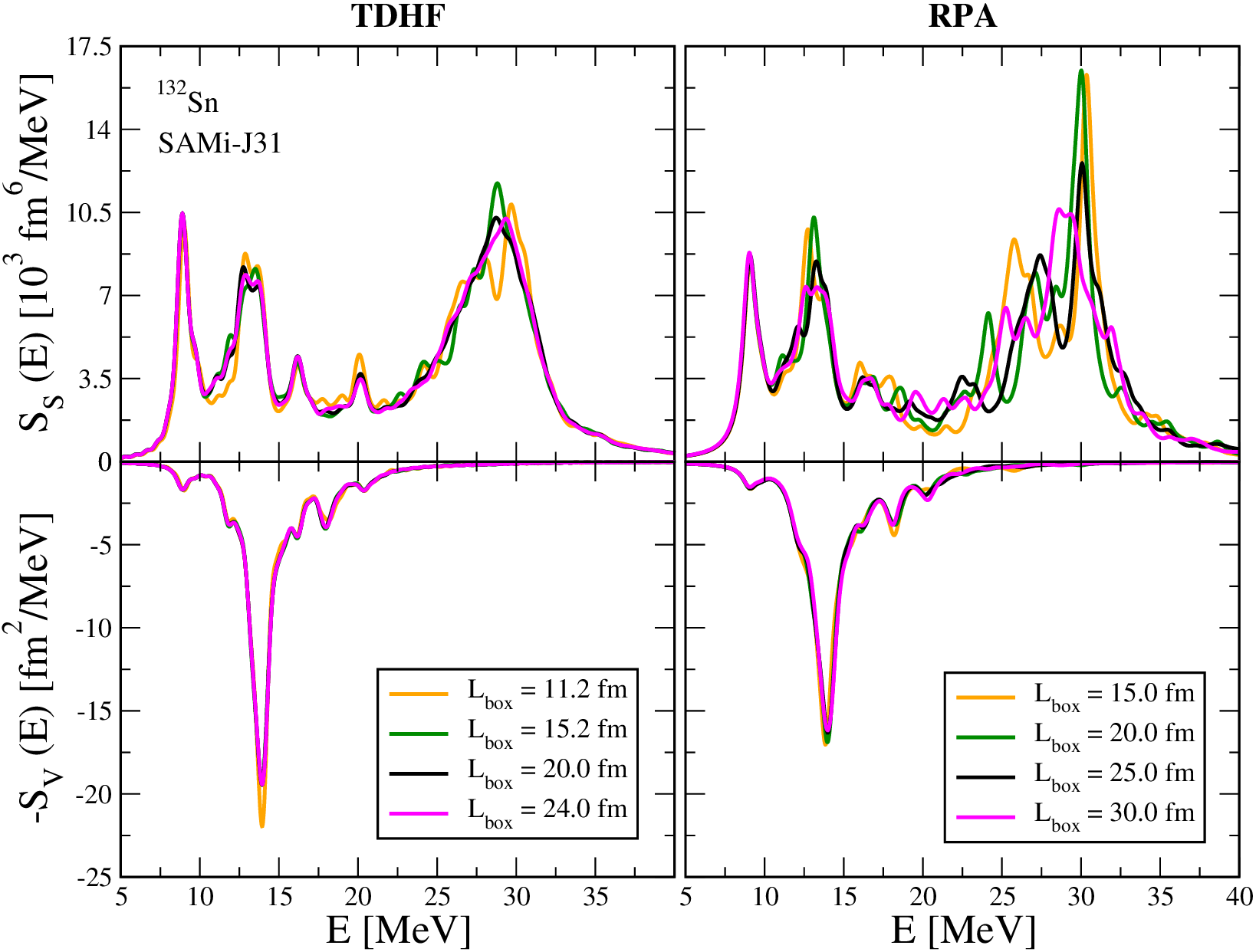} 
\caption{(Color online) The strength function of the IS (top panels) and IV (bottom panels) response as obtained in TDHF (left panels) or in RPA calculation (right panels) for $^{132}$Sn with SAMi-J31 interaction with different box sizes.}
\label{comparison_new}
\end{figure}

For this analysis, we discuss the results obtained for the  $^{132}$Sn system,
employing the SAMi-J31 interaction.  
Fig.~\ref{comparison_new} presents a comparison of the IS and IV dipole response, as obtained in TDHF calculations, with corresponding RPA calculations. 
In the latter, the strength function is calculated by convoluting the transition probability with a Lorentzian function of width equal to 0.5 MeV. 
On the other hand, in the TDHF results, the spread originates from the finite time interval considered to follow the dynamics and from our smoothing 
procedure (see Section~\ref{sec:dipole}).

For the two approaches, we have explored the dependence of the results on the size of the box considered.  This ingredient determines the {discretization 
of the single-particle states in the continuum, so that it could affect the details of the oscillation modes at higher frequency}.
We denote by $L_{box}$ either half of the side of the cubic box employed in THDF
calculations or the radius of the spherical box considered in RPA calculations.
Within the spanned range, the values obtained for both the IS and IV
Energy Weighted Sum Rule (EWSR) are convergent 
and consistent between 
THDF and RPA calculations 
(within 0.1$\%$). 

As one can see in Fig.~\ref{comparison_new}, the IV response is
nearly insensitive to  $L_{box}$  (within the range considered) and
an excellent agreement is obtained between the two approaches. {The excitation energies of the modes {characterizing} the IV response are indeed 
{lower, and the coupling with the continuum is smaller than 
in the case of the IS response.} 
As far as the IS component is concerned, the TDHF response is slightly affected by the $L_{box}$ parameter and practically converges to its final shape already for $L_{box}$ = 20 fm.  
{The RPA calculations exhibit, within a similar range of values as adopted in 
the TDHF case, a larger sensitivity to $L_{box}$ in the 
high energy region of the IS spectrum. 
At present, although we cannot prove it, we could say that the
differences between TDHF and RPA 
may be simply due to different discretization procedures. 
An indication
along this line can be seen in Fig.~\ref{comparison_sn132}, that shows
a comparison between TDHF ($L_{box} =$ 20 fm) 
and RPA ($L_{box} =$ 30 fm).}
Here, {at variance with Fig.~\ref{comparison_new},} 
the RPA strength has been convoluted 
with a Lorentzian function of energy-dependent width
\begin{equation} 
\Gamma (E) = 
\begin{cases}
e^{-\frac{(E-30)^2}{25^2} \ln 2} \qquad 5 \le E \le 55 \,\, \textup{MeV}\\
0.5 \qquad \qquad \qquad \textup{elsewhere,}
\end{cases}  
\end{equation}
{which leads to a maximum width of 1.0 MeV in the energy region of the IS GDR.}

\begin{figure}[t]
\includegraphics[scale=0.3]{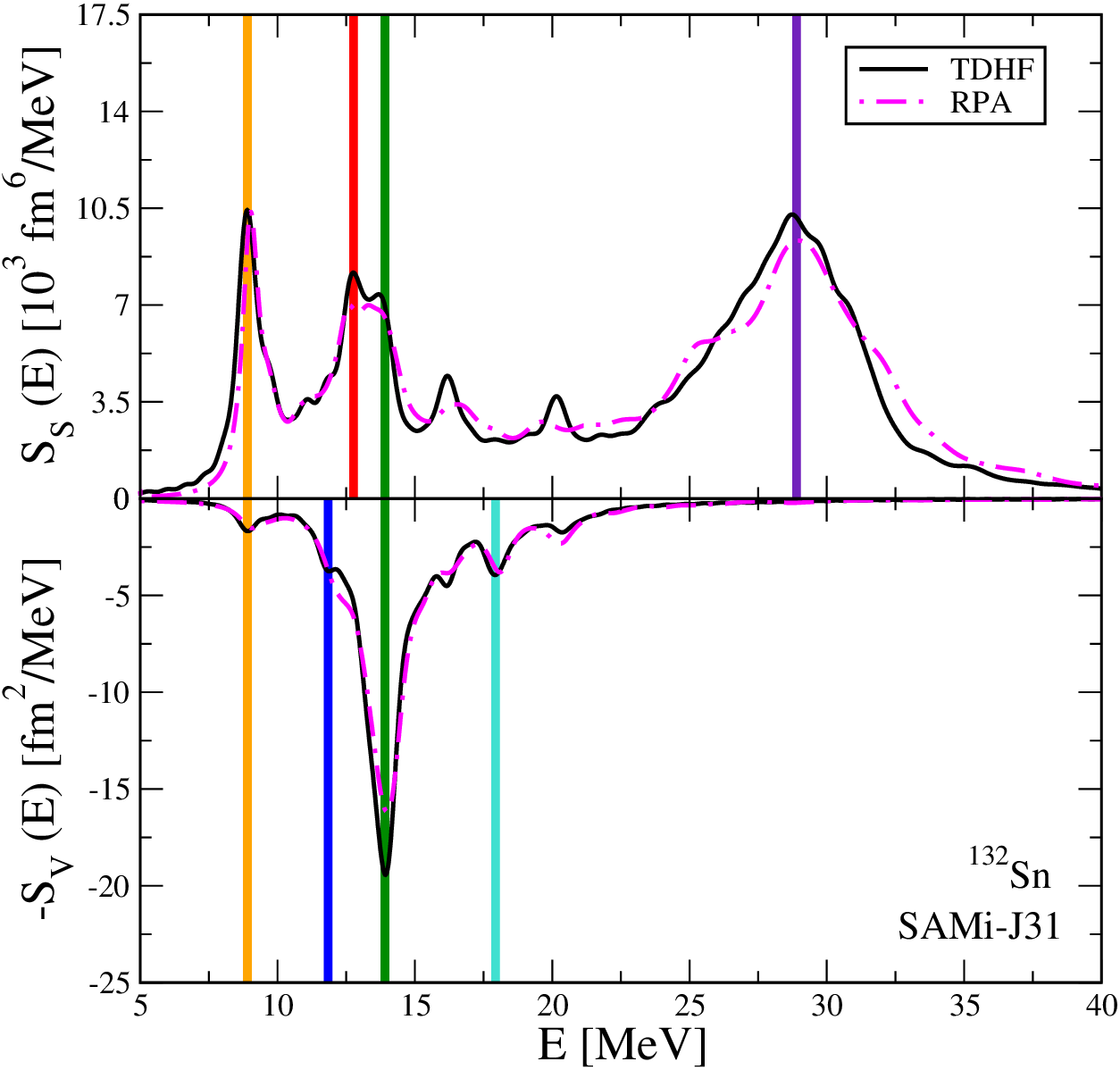}
\caption{(Color online) The strength function of the IS (top panel) and IV (bottom panel) response for $^{132}$Sn with SAMi-J31 as obtained in TDHF or in RPA calculation, {with $L_{box} =$ 20 or 30 fm, respectively}. 
The vertical lines indicate the energy of the modes selected for the transition
density analysis (see Section III.E).}
\label{comparison_sn132}
\end{figure}

One can conclude that, in spite of the different degree of sensitivity to some
technical ingredients, such as the box size, a very good agreement is observed
between TDHF and RPA calculations, as far as the IS and IV dipole responses are concerned.  






\subsection{Transition densities}
In addition to the investigation of the dipole strength discussed so far, the analysis of the transition densities associated with the different excitation modes of the system is very instructive since it delivers important information about the spatial structure related to the dynamics of every excitation. To undertake this analysis in TDHF and Vlasov calculations, we need to evaluate the local spatial density as a function of time. In order to reduce numerical fluctuations, we take into account the cylindrical symmetry of the initial perturbation  and, averaging over the azimuthal $\phi$ angle, we extract the density $\rho_q (r, \cos \theta,t)$ and the corresponding fluctuation $\delta \rho_q (r, \cos \theta,t) = \rho_q (r, \cos \theta, t)$ - $\rho_q (r, t_0)$, where $\cos \theta = z/r$ and $\rho_q (r, t_0)$ denotes the ground state density profile, which only depends on $r$. As suggested in Ref.~\cite{urbPRC2012}, assuming that the amplitude of the oscillation is weak (linear response regime), the spherical symmetry of the ground state and the dipole shape of the excitation operator imply that the transition density can be written, at each time, as $\delta \rho_q (r, \cos \theta, t) = \delta \rho_q (r,t) \cos \theta$. Then one can finally extract the transition density just as a function of the radial distance $r$, by averaging over the polar angle
the quantity $\delta \rho_q (r,t)$. 

\begin{figure}[b]
\includegraphics[scale=0.3]{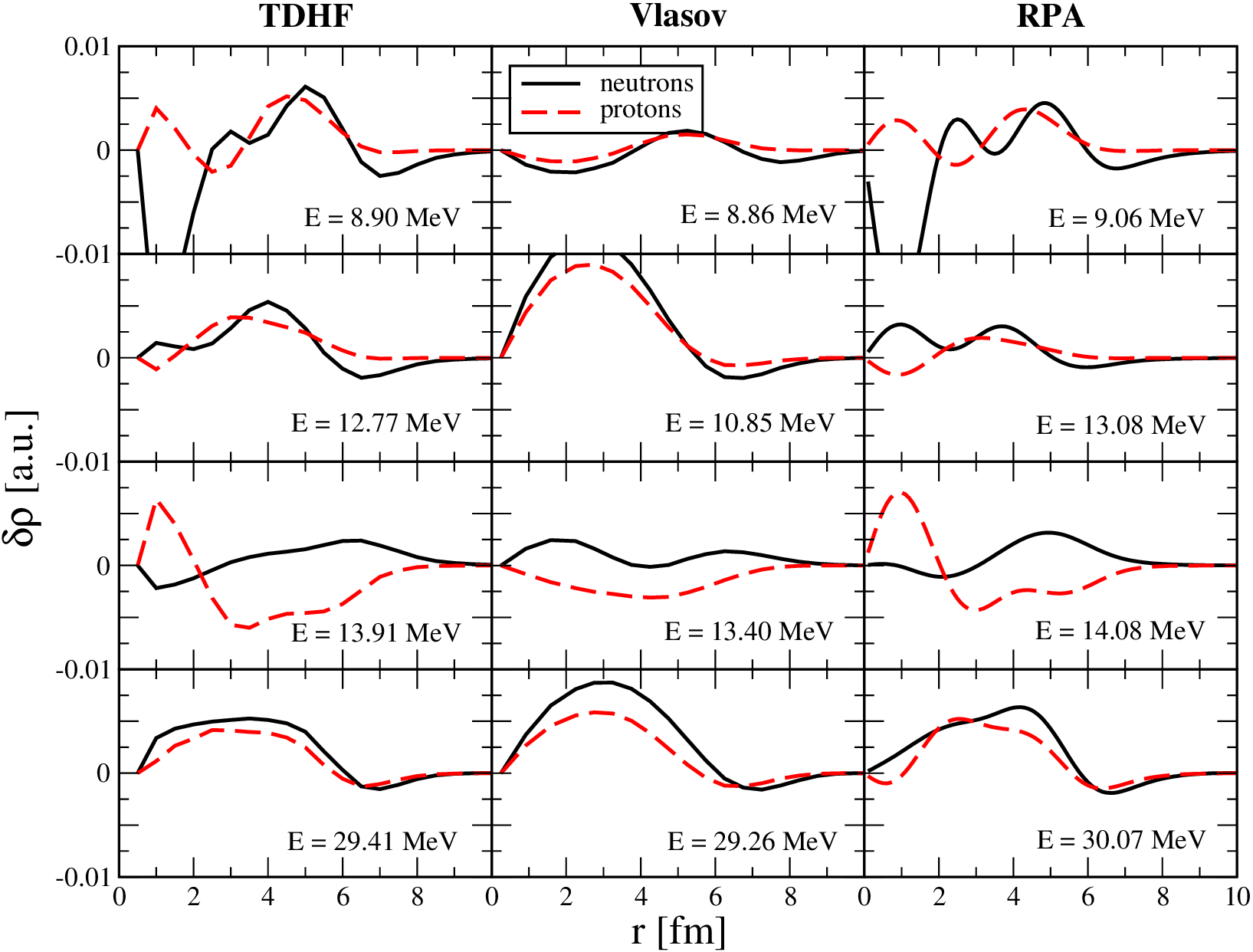}
\caption{(Color online) The transition densities related to the main peaks evidenced in Fig.~\ref{comparison_sn132}, as obtained in the IS strength function for $^{132}$Sn, SAMi-J31 parameterization and for the three different models considered. Corresponding energies are also indicated.}
\label{fig04} 
\end{figure}
It is clear that, both in Vlasov and TDHF calculations, the perturbation V$_{ext}$, at $t=t_0$ , {induces} simultaneously all modes which can be excited by the operator $\hat{D}_k$. Thus the corresponding density oscillations observed along the dynamical evolution will appear as the result of the combination of the different excitation modes. 
In order to pin down the contribution of a given mode, of energy $E$, to the density oscillations, one can 
compute the Fourier transform of $\delta \rho_q (r,t)$:
\begin{equation}
\delta \rho_q (r,E) \propto \int_{t_0}^\infty dt~\delta \rho_q (r, t) \sin \frac{E t}{\hbar}.
\end{equation}
In practice, since the simulation runs only to $t_{max} = 1800$ fm/c, the sine function is multiplied by a damping factor, as in the strength function $S_k$ (E). 

We notice that, in RPA calculations, one does not need to use any auxiliary 
prescription, since the transition densities are directly evaluated from the forward and backward amplitudes solution of the 
RPA matrix, associated with a given energy {eigen}value $E$ [see Eqs.~(36)-(37) in Ref.~\cite{rpa2013}]. {Nonetheless, in principle, it could be
possible to average the RPA transition densities in a given energy window.} 

It is well known that, in symmetric matter, neutrons and protons oscillate with exactly equal (isoscalar) or opposite (isovector) amplitudes. In neutron-rich systems, the picture is more complex; however, one can still identify isoscalar-like modes, when the two nuclear species oscillate in phase, and isovector-like modes, with neutrons and protons oscillating out of phase. Apart from this information, connected to the mixed character of each mode, the overall spatial structure of the transition densities tells us which part of the system 
(internal part or surface) is more involved in the oscillation.

\begin{figure}[b]
\includegraphics[scale=0.3]{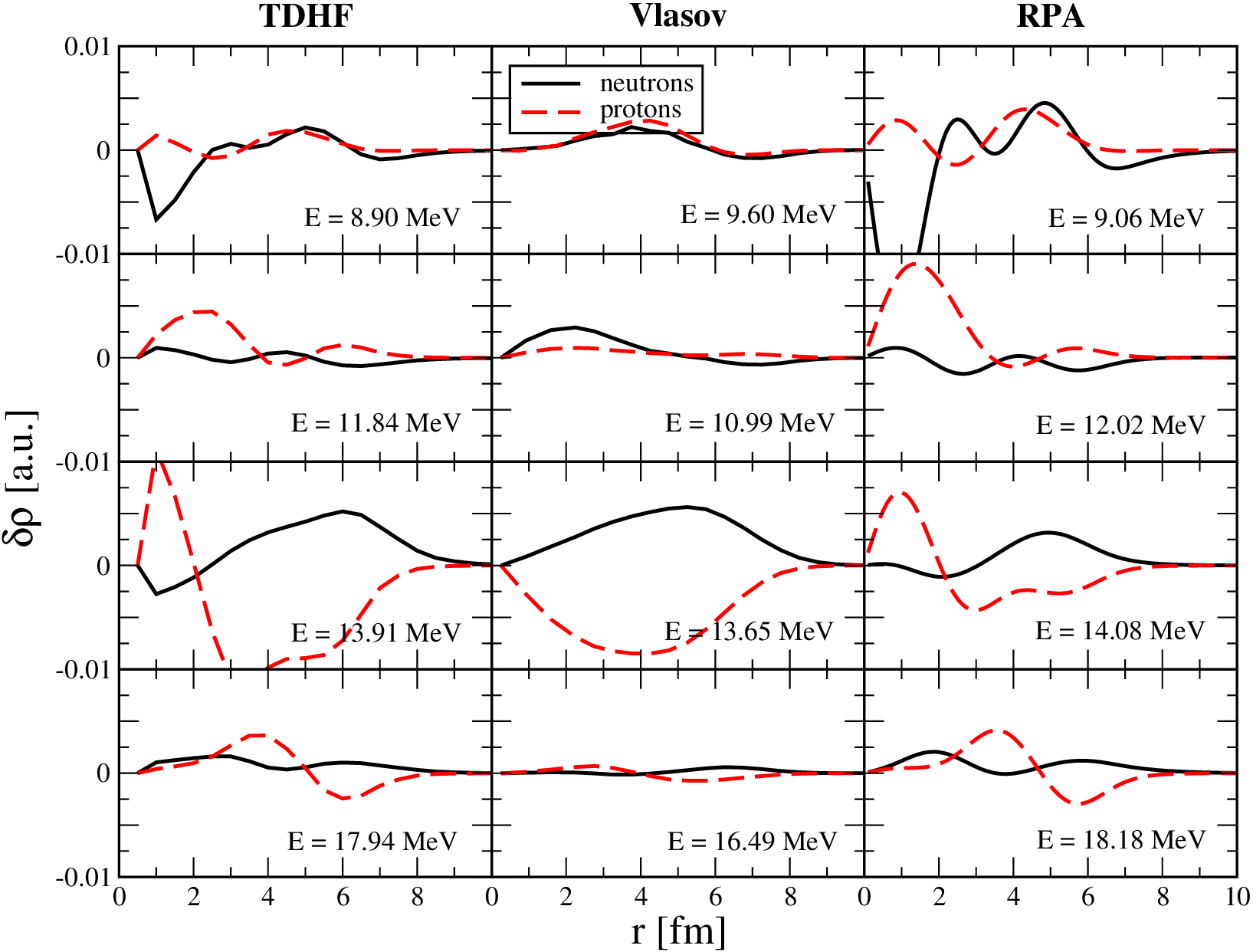}
\caption{(Color online) The transition densities related to the main peaks evidenced in Fig.~\ref{comparison_sn132}, as obtained in the IV strength function for $^{132}$Sn, SAMi-J31 parameterization and for the three different models considered. Corresponding energies are also indicated.}
\label{fig05} 
\end{figure}
In dynamical calculations, dipole excitations are directly excited by a 
given (IS or IV) perturbation. 
Hence IS(IV)-like oscillations, and corresponding transition densities,  
are better identified when an initial IS(IV) perturbation is applied. 
Actually, this possibility to directly probe the response of the system to
specific excitations could also help to disentangle between modes having similar energies but different nature. 
On the other hand, the modes with a strong IS/IV mixing react to both
(IS and IV) excitations, so 
the associated transition density can be extracted 
from both kinds of calculations.       

Here we present the transition densities related to the modes giving a 
sizeable contribution to the IS dipole strength function (Fig.~\ref{fig04}) 
and/or to the IV one (Fig.~\ref{fig05}), as obtained for the system $^{132}$Sn
in TDHF, Vlasov and RPA calculations. 
For the Vlasov calculations, we consider the same modes identified in 
Ref.~\cite{zhePRC2016}.
The energies considered in TDHF and RPA calculations are indicated {by vertical bars} in 
Fig.~\ref{comparison_sn132}.

As a general feature, it should be noticed that TDHF and RPA calculations lead to very similar results. 
The first row of the two figures displays the structure of 
{what we may call PDR} (orange bar in Fig.~\ref{comparison_sn132}), 
which manifests {itself} as an isoscalar-like mode, but with also 
an isovector contribution. 
Indeed, in TDHF and Vlasov calculations,  
essentially the same structure is observed when the transition density
is extracted from IS (left panel) or IV (right panel) perturbations, though 
with a reduced amplitude in the latter case.  

The structure obtained in quantal calculations is in agreement with previous results~\cite{mazPRC2012} and is  {qualitatively} well reproduced also by the semi-classical density oscillations, except for the behavior in the central region which could be related to the trend observed in the quantal isovector density profiles (see Fig.~\ref{fig02}).  One can see that density oscillations
involve deeply the surface region (see the behavior for 
$r$ between 5 and 9 fm). 
This is in line with the observation that this mode is particularly robust in
nuclei exhibiting a diffuse density profile, as discussed in Section~\ref{sec:comparison_classical_quantal}. 

In the low-lying energy domain (below the IV GDR), a second peak is observed 
both in the IS and IV dipole strength (blue and red bars in 
Fig.~\ref{comparison_sn132}, respectively).  These peaks reflect two distinct
excitation modes, though their energy is close.  The corresponding 
transition densities 
are displayed in the second row of Fig.~\ref{fig04} and Fig.~\ref{fig05} and, 
as it is particularly clear in the Vlasov case, manifest 
their isoscalar-like or isovector-like nature, respectively. This result indicates that, in addition to the PDR,  it is possible to recognize {at least} other two distinct modes, with different structure, in the energy region below the IV GDR.
As compared to the PDR, for this second isoscalar-like mode (compare first and second rows of Fig.~\ref{fig04}) density oscillations look shifted to the left, 
thus involving more the internal part of the system. Actually, this mode
should correspond to standard toroidal excitations, where the surface
moves against the core.  Thus, we expect this mode to be more robust in nuclei
with a sharper evolution from the volume to the surface in the density profile.
   
{A deeper investigation on the nature of the modes lying in this low-energy domain will be tackled in Sec.~\ref{sec:low_lying_modes}.}

The structure of the IV GDR (green bar in Fig.~\ref{comparison_sn132}) is plotted in the third rows of the two figures. 
In all the cases, the well-established isovector-like structure corresponding to the semi-classical Goldhaber-Teller (GT) picture is well represented, with essentially one prominent oscillation, having a maximum close to the nuclear 
surface (see Fig.~\ref{fig05})~\cite{goldhaber1948}. The mode presents also a 
sizeable isoscalar component; indeed quite similar transition densities, 
though of reduced amplitude, are
extracted considering an initial IS perturbation (Fig.~\ref{fig04})~\cite{stein1950}. 

Lastly, the last rows are dedicated to display the structure of two volume modes: the IS GDR peak obtained in the high energy region of the IS response (Fig.~\ref{fig04}, violet bar in Fig.\ref{comparison_sn132}) and the isovector-like peak emerging in the IV response beyond the IV GDR (Fig.~\ref{fig05}, cyan bar in Fig.~\ref{comparison_sn132}). 
One can notice that the latter IV peak exhibits a structure which is typical of the Steinwedel-Jensen (SJ) description, characterized by a kind of double oscillation and deeply involving also the internal part of the system.
The three models compare very well in this case. 

To conclude, from this analysis it emerges that semi-classical calculations
are able to grasp the main features of the density oscillations associated
with the excitation modes considered here, though volume modes
are described better than the ones characterized by important surface contributions. 

 


\subsection{Low-lying energy modes for Sn isotopes}
\label{sec:low_lying_modes}
We focus here on low-energy modes, which are more intriguing and controversial, 
exploring how their features evolve with the isospin asymmetry content of the systems.
A large amount of investigations has been devoted in the last years to the behavior of 
a variety of isotopes, 
from light to heavy, from spherical to deformed, and from normal to superfluid nuclei, in order to shed light on the properties of the PDR~\cite{ebaPRC2014}. 
Although we are not going to develop here a systematic study in the strict sense, with the aim of elucidating our understanding of the structure of the low-lying energy modes, in this section we are looking at the properties of these excitations in three spherical nuclei belonging to the Sn isotope chain: 
{the semi-magic nucleus $^{120}$Sn and two double magic-nuclei, namely $^{100}$Sn and $^{132}$Sn. The latter Sn isotope is the one already considered in the 
previous
sections.} In such a way, we can isolate the effect of the N/Z ratio on the isoscalar/isovector mixing and on the structure of the modes we wish to analyze. 

\begin{figure}[t]
\includegraphics[scale=0.3]{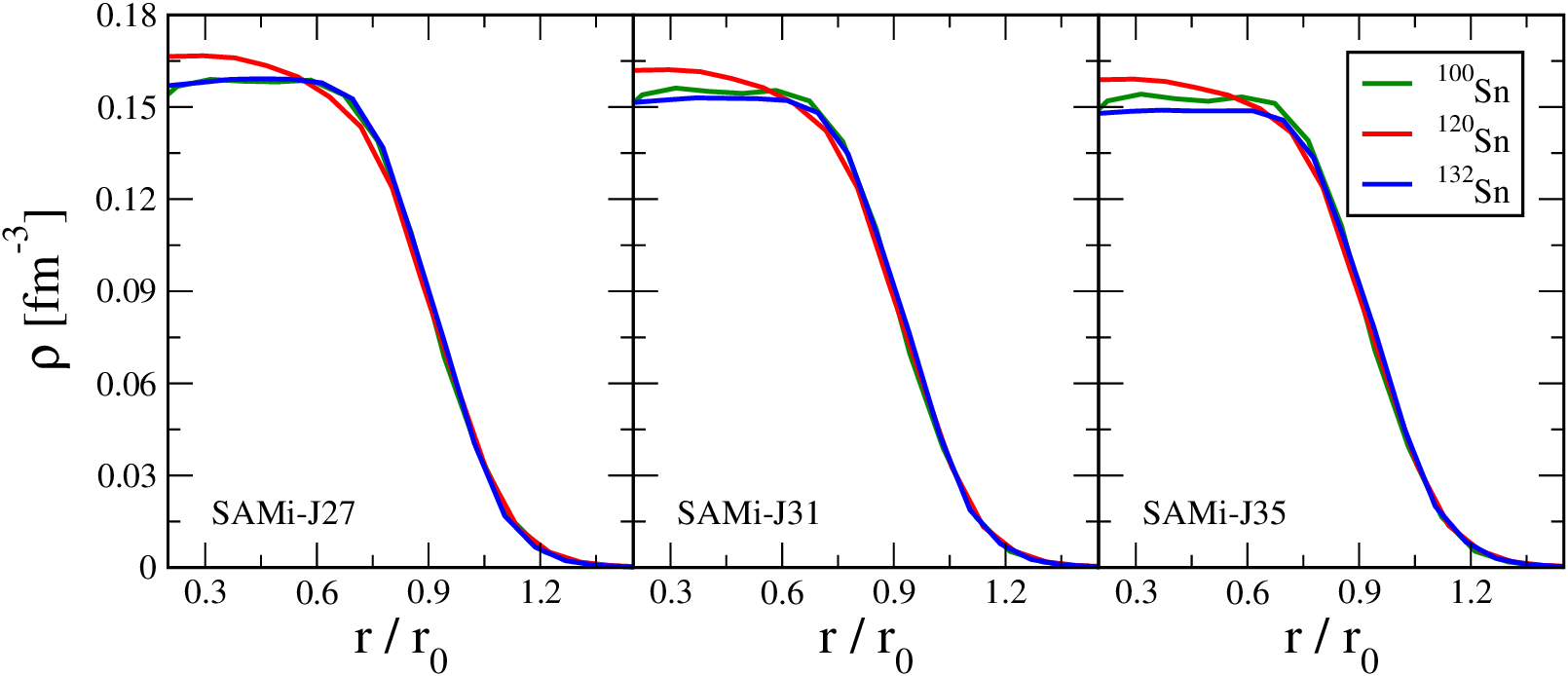}
\caption{(Color online) The isoscalar density profiles 
of the three spherical nuclei in the Sn isotope chain, for the three parametrizations SAMi-J27, SAMi-J31 and SAMi-J35 considered ($r_0 = 1.2 A^{1/3}$).}
\label{fig19} 
\end{figure}

In analogy with the investigation carried on for the closed shell nuclei examined in the first section, in Figs.~\ref{fig19} - \ref{fig20} the isoscalar density and {local} asymmetry profiles 
are plotted, for the three Sn isotopes considered. The three 
SAMi-J parameterizations of the effective interaction introduced above
are adopted: SAMi-J27, SAMi-J31, SAMi-J35. In such a way, it will also 
be possible to probe the effects of modifying the isovector channel 
{of the functional considered} on the observables under study. In order to better compare the structure of these profiles, in these figures, as in Figs.~\ref{fig03} and~\ref{fig02}, we renormalized the radius with respect to the standard radius $r_0$ ($r_0= 1.2 A^{1/3}$). 
From Fig.~\ref{fig19}, it is rather evident that, as a consequence of the shell structure, the double magic nuclei exhibit a similar profile, which has a more compact shape and a rather flat behavior in the internal region. This configuration reflects in a sharper radial evolution of the density in the surface region, with respect to the open-shell nucleus $^{120}$Sn, whose density profile appears more diffuse. 
At the same time, the isovector density profiles (Fig.~\ref{fig20}) 
clearly show the increasing of the {local} asymmetry 
$\rho_3 / \rho$ in correspondence of the surface, especially for the more neutron-rich system, $^{132}$Sn, owing to the neutron skin development~\cite{TrzPRL2001}.
This is more evident employing the SAMi-J35 interaction, that has the largest value of the slope L. 
One can also observe a sligthly proton-rich surface region in the case of 
 $^{100}$Sn.


\begin{figure}[b]
\includegraphics[scale=0.3]{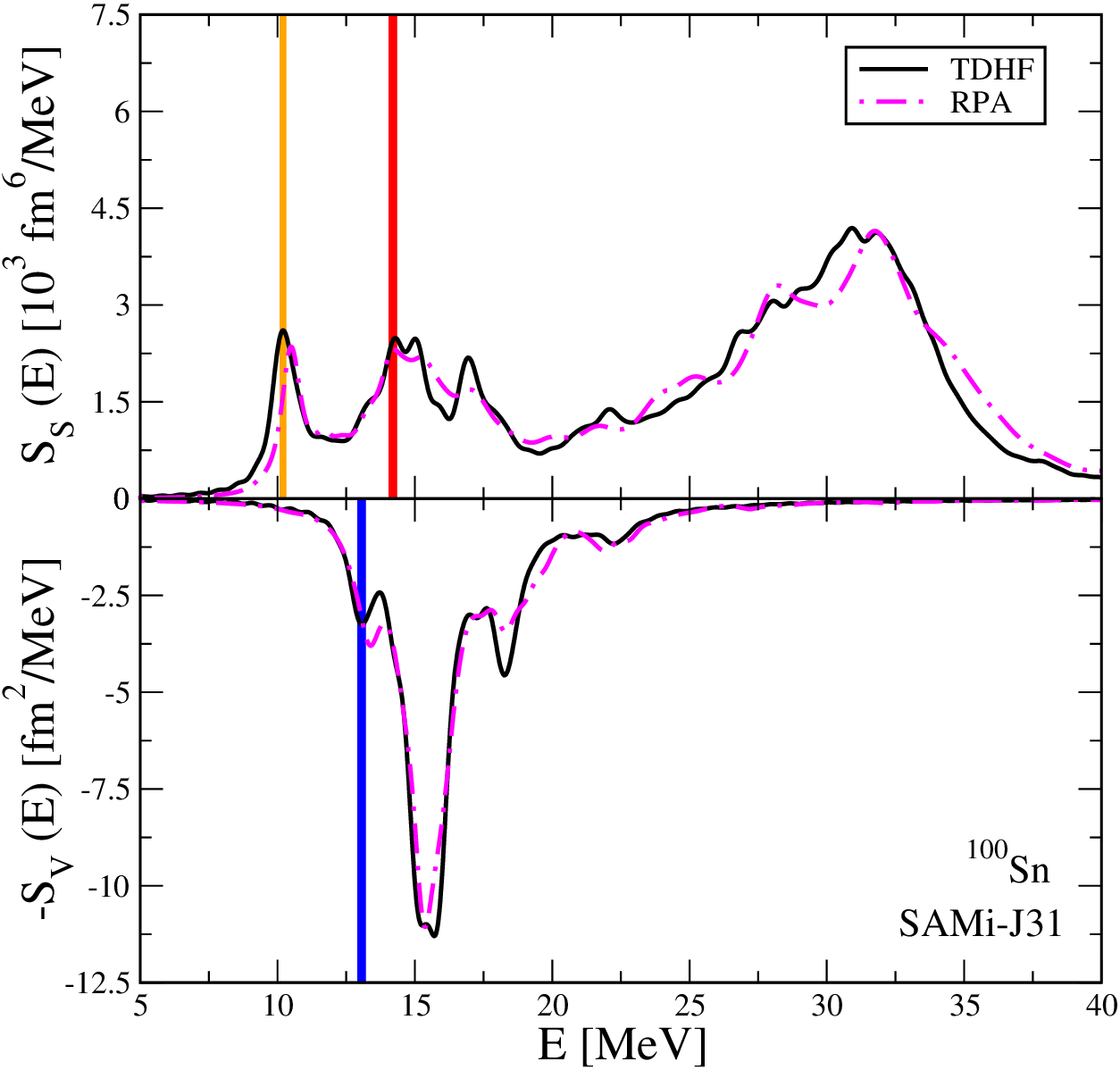}
\caption{(Color online) Similar to Fig.~\ref{comparison_sn132}, but for $^{100}$Sn.}
\label{fig08} 
\end{figure}


Analogously to the analysis presented in Fig.~\ref{comparison_sn132} for $^{132}$Sn, in Figs.~\ref{fig08} - \ref{fig09}, we show the IS and IV dipole strengths as obtained with both TDHF and RPA models, for {$^{100}$Sn and} $^{120}$Sn considering the SAMi-J31 parameterization only. 
We observe a nice agreement between TDHF and RPA calculations. 
It clearly emerges that the IS dipole strength of the pygmy mode (indicated by the orange bar) is strongly enhanced in the case of $^{120}$Sn, not only with respect to the symmetric system $^{100}$Sn, but also in comparison to the neutron-rich nucleus $^{132}$Sn. The relative importance of the pygmy mode in the IS response is in fact enforced at the expense of the strength arising in the energy region just  below the IV GDR {(peak associated with the red bars in Figs.~\ref{comparison_sn132}, \ref{fig08} and \ref{fig09}).} 
As already discussed in Section~\ref{sec:comparison_classical_quantal}, when commenting the differences observed between Vlasov and TDHF calculations, this evolution could be connected to the different density profile of the open-shell $^{120}$Sn, with respect to the closed shell nuclei. 
Also here, one can notice that a smoother density profile
is associated with a larger strength of the isoscalar mode {of lowest energy}, that has a significant surface component.
Owing to the coupling which exists in isospin asymmetric systems, 
 {a larger PDS is observed in the IV response. In other words, the PDS does not increase monotonically with N, contrarily to the trend exhibited by the neutron-skin in these Sn isotopes.}   

\begin{figure}[t]
\includegraphics[scale=0.3]{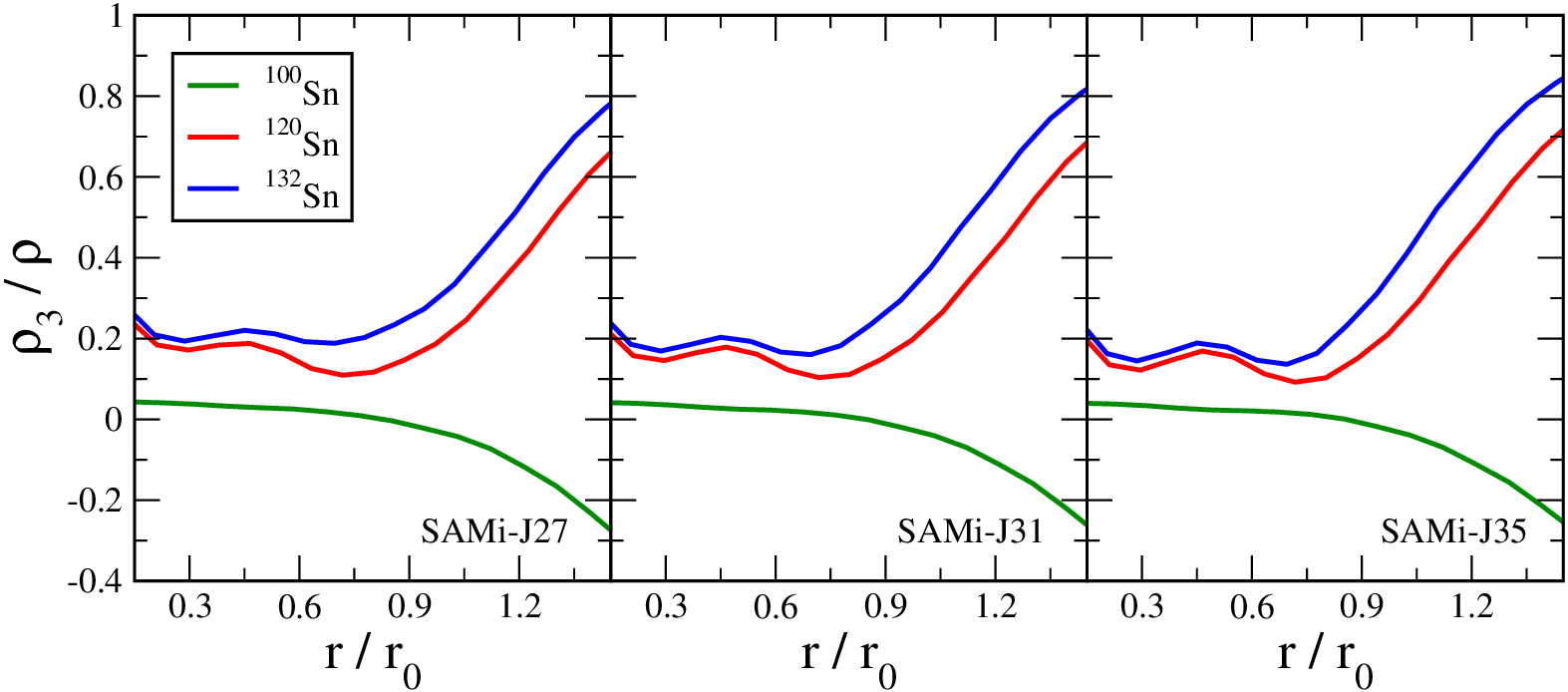}
\caption{(Color online) The {local} asymmetry profiles 
of the three spherical nuclei in the Sn isotope chain, for the three parametrizations SAMi-J27, SAMi-J31 and SAMi-J35 considered ($r_0 = 1.2 A^{1/3}$).}
\label{fig20} 
\end{figure}

\begin{figure}[b]
\includegraphics[scale=0.3]{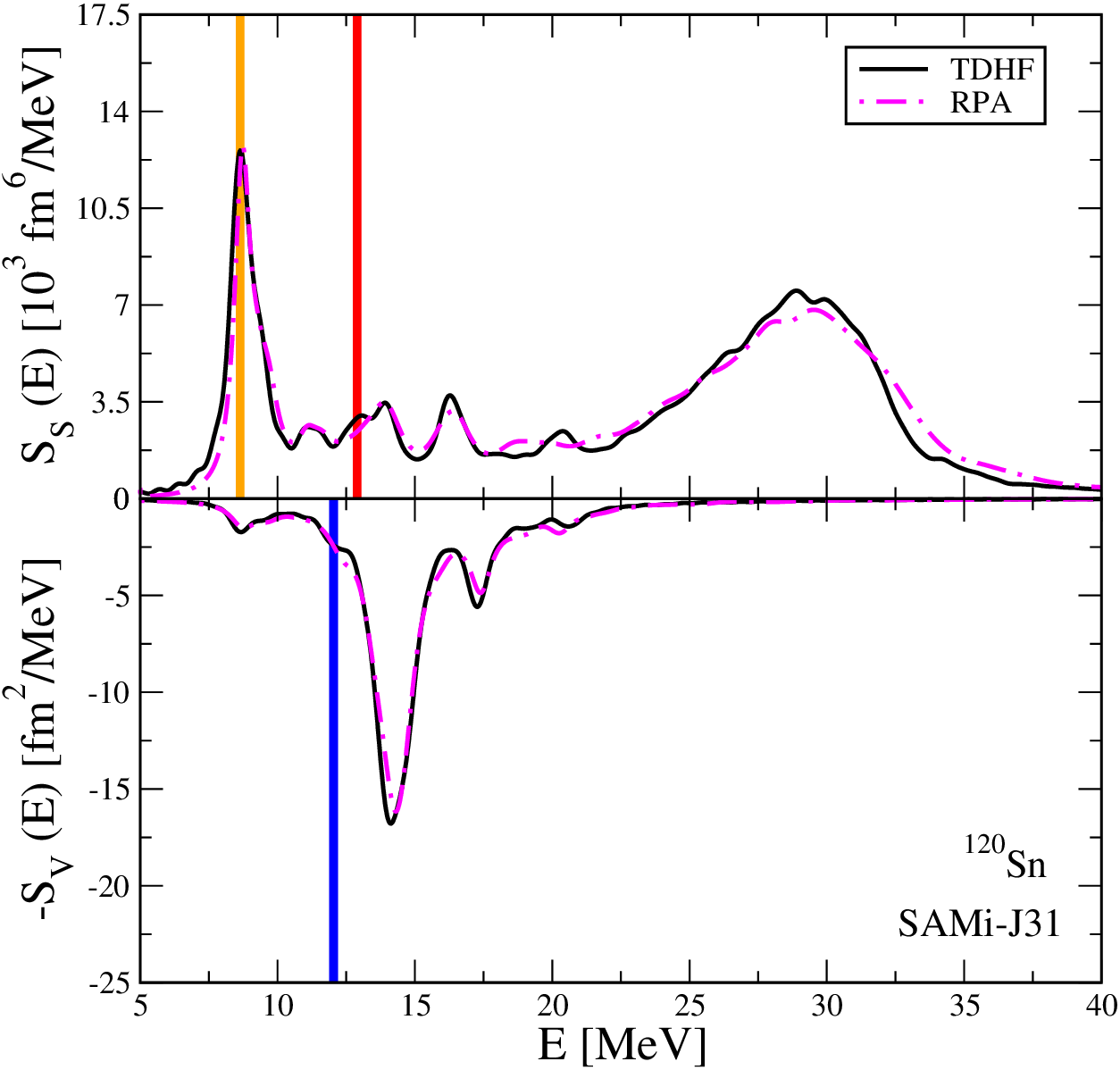}
\caption{(Color online) Similar to Fig.~\ref{comparison_sn132}, but for $^{120}$Sn.}
\label{fig09} 
\end{figure}


\begin{figure*}[t]
\includegraphics[scale=0.3]{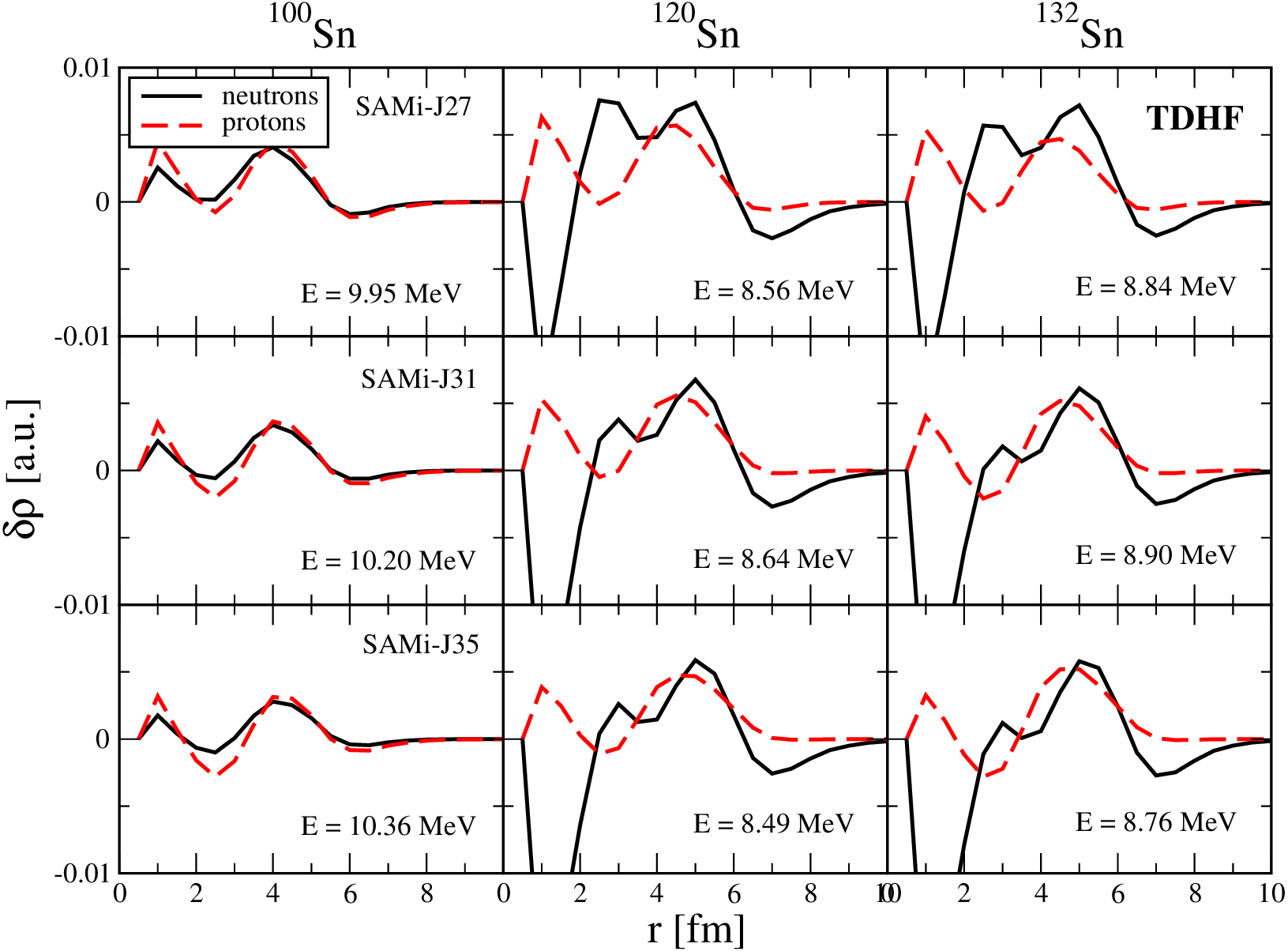} \qquad 
\includegraphics[scale=0.3]{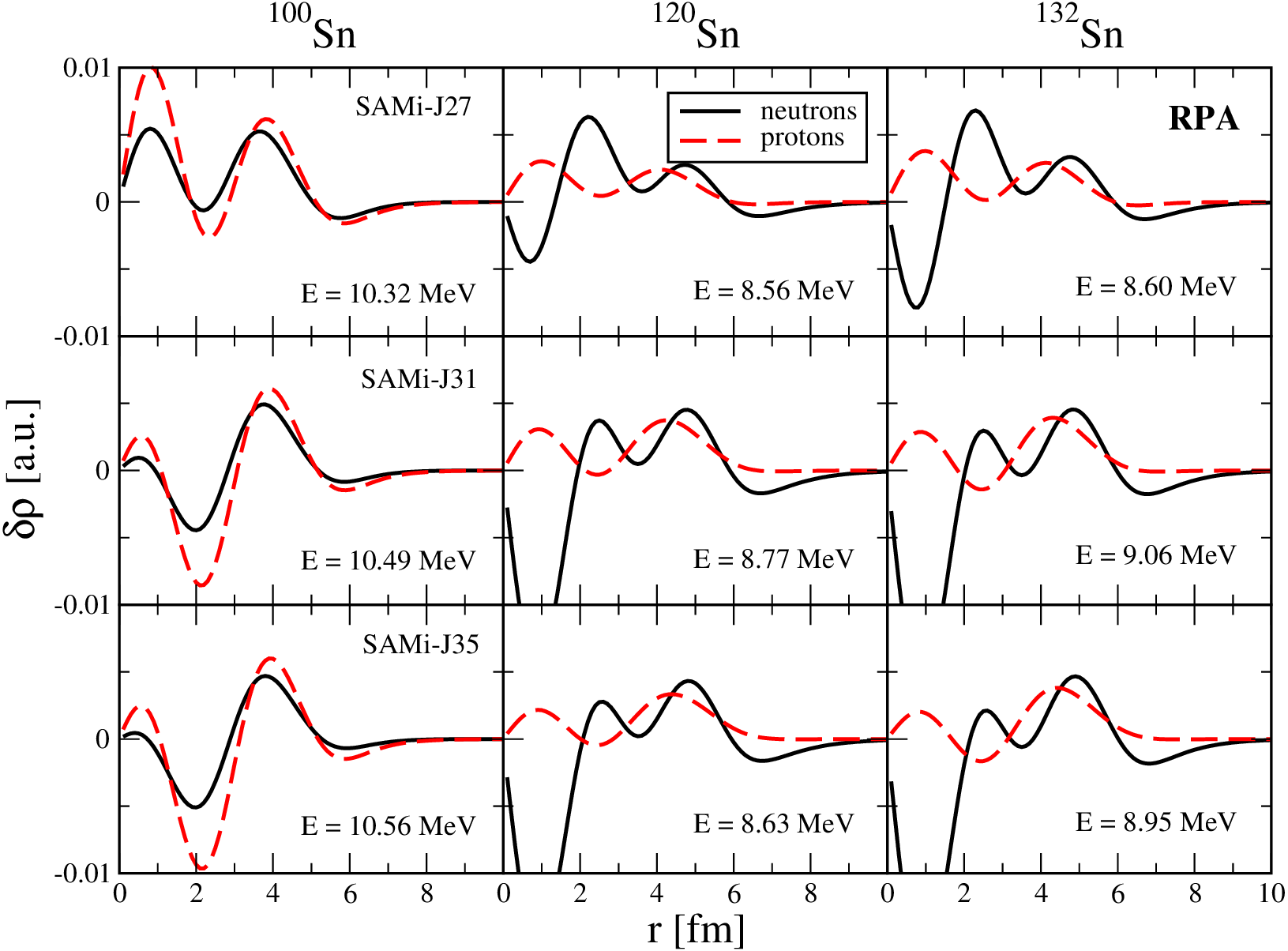}
\caption{(Color online) The transition densities of the PDR, as obtained in TDHF with an initial IS perturbation (left panels) or in RPA (right panels) calculations, for the three nuclei in the Sn isotope chain and for the three SAMi-J parameterizations.}
\label{fig13} 
\end{figure*}


Let us concentrate now on the spatial structure of the low-lying energy modes.
In the following we will investigate the Sn isotopes introduced above 
and we will present the results for the three parameterizations of the effective interaction employed in our study. The {left panels} in Fig.~\ref{fig13} present the transition densities extracted in TDHF for the lowest energy peak in the pygmy region of the IS dipole response {(indicated by orange bars in Figs.~\ref{comparison_sn132}, \ref{fig08} and \ref{fig09})}. 

\begin{figure}[h]
\includegraphics[scale=0.3]{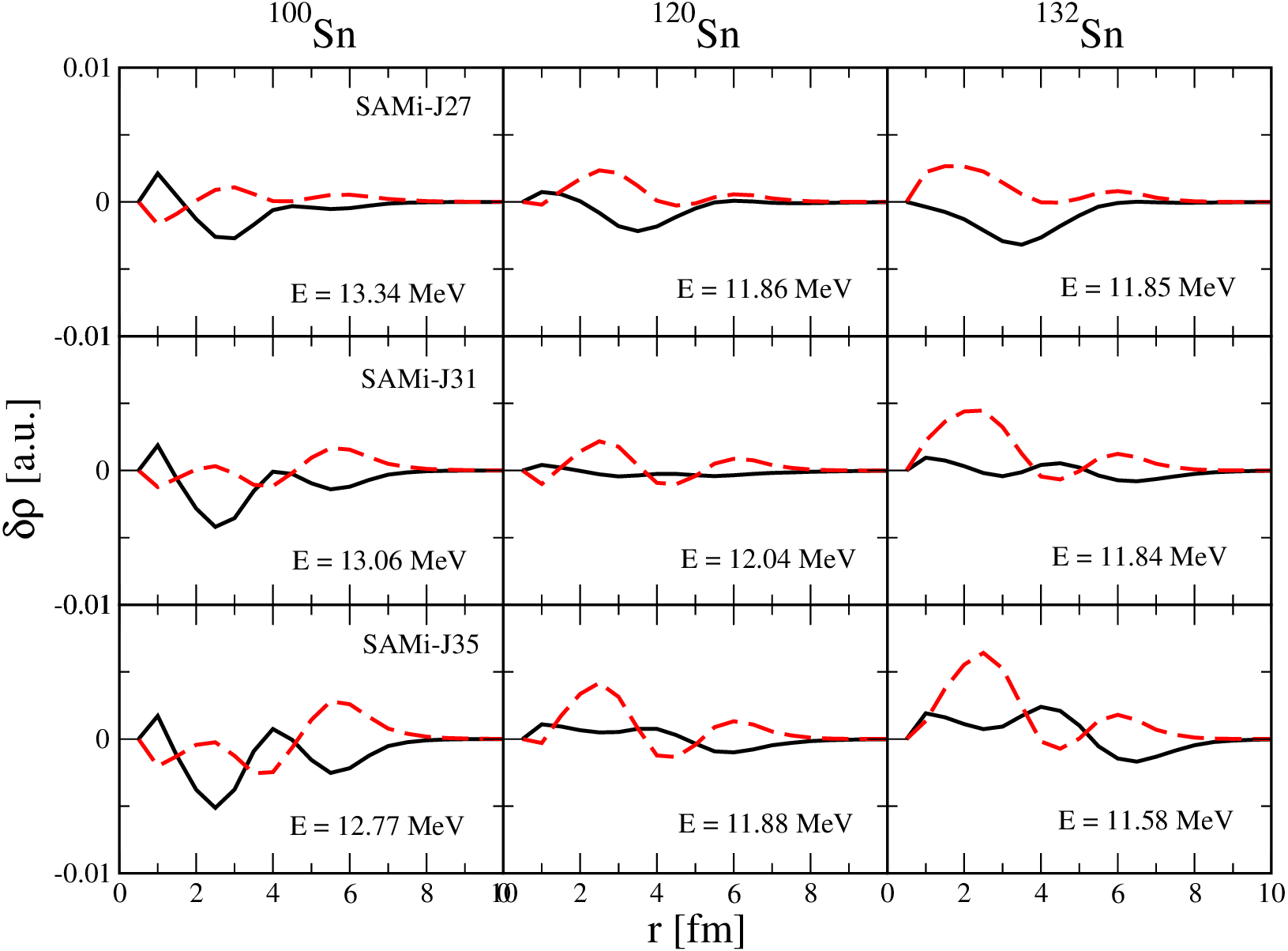} 
\caption{(Color online) The transition densities of the second IV peak (indicated by blue bars in Figs.~\ref{comparison_sn132}, \ref{fig08} and \ref{fig09}), for the three Sn isotopes. The results are obtained in TDHF and for three SAMi-J parameterizations.}
\label{fig16} 
\end{figure}

\noindent The result related to the system $^{132}$Sn and the SAMi-J31 interaction has already been shown in Fig.~\ref{fig04}. 
However, here our aim is to see the evolution of the PDR structure when varying the isospin asymmetry of the systems, as well as the interaction adopted. First of all, it is interesting to notice once again the isoscalar-like nature of the PDR, especially for the symmetric system, where neutrons and protons oscillate almost exactly in phase. As extensively discussed above, indeed, for this system the isoscalar/isovector mixing{, which usually characterizes neutron-rich systems,} is strongly reduced. 
Moreover, in agreement also with our semi-classical results~\cite{zhePRC2016}, when considering interactions with increasing slope L (from SAMi- J27 to SAMi-J35), one can see that {for neutron-rich systems,} neutron oscillations become larger, with respect to proton oscillations, especially in the surface region. This can be explained on the basis of Fig.~\ref{fig20}, where one observes that the system asymmetry is pushed more towards the surface, corresponding to the development of a thicker neutron skin, when increasing the value of the slope L.

\begin{figure}
\includegraphics[scale=0.3]{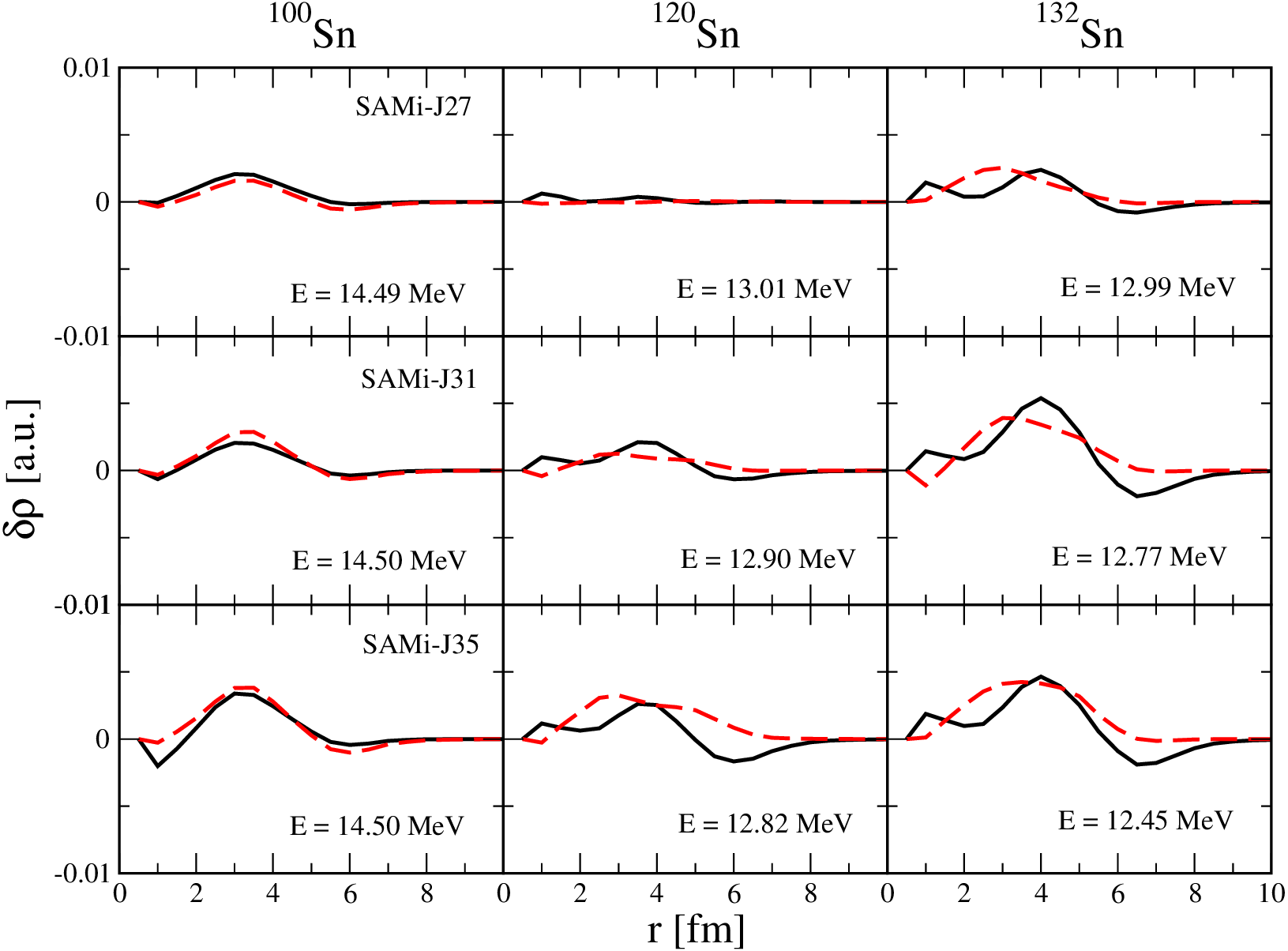}
\caption{(Color online) The transition densities of the second IS peak (red bars in Figs.~\ref{comparison_sn132}, \ref{fig08} and \ref{fig09}), for the three Sn isotopes. The results are obtained in TDHF  and for three SAMi-J parameterizations.}
\label{fig17} 
\end{figure}

The right panels of Fig.~\ref{fig13} show that a good agreement is obtained
with the analogous RPA results.



We turn now to examine the peak observed in the IV response (the blue bars
in  Figs.~\ref{comparison_sn132}, \ref{fig08} and \ref{fig09}).

As shown by the TDHF calculations of Fig.~\ref{fig16}, 
the transition densities clearly 
reveal the isovector-like nature of this mode. Again, as for the PDR, the amplitude of the oscillations at the surface increases as a function of the slope L 
{(see Fig.~\ref{comparison_TDHF_samiall} for the corresponding effect on the strength)}.
Moreover, the IS/IV mixing increases when moving from $^{100}$Sn to  $^{132}$Sn.
It is interesting to notice that the observation of two low-energy modes of close energy, the one of lowest energy being mainly isoscalar and the other being
mainly isovector,  
has been reported also in other recent studies \cite{braEPJA2015}.  
 

Finally, Fig.~\ref{fig17} shows the transition densities of the second relevant
peak appearing in the IS response 
{(that is the peak indicated by red bars in Figs.~\ref{comparison_sn132}, \ref{fig08} and \ref{fig09})}.

As discussed above,  
this excitation {may correspond to the toroidal mode}~\cite{urbPRC2012}. 
Indeed, the transition densities deduced by employing an IS perturbation manifest the development of a mode which is clearly isoscalar-like, especially in the case of $^{100}$Sn where coupling effects are 
{quenched} (see Fig.~\ref{fig17}). 
Also in this case, the IS/IV mixing increases with the symmetry energy 
slope L of the parametrization considered and with the N/Z of the system. 
The shape of the transition densities indicates that
this mode corresponds to oscillations of the surface against the volume. 
The small amplitude observed for $^{120}$Sn, especially in the case of SAMi-J27, is related to the reduced IS strength in the energy region considered (see Fig.~\ref{fig09}).  It should be noticed that a quite good correspondence
with RPA results {(not shown here)} is obtained also for the modes described in 
Figs.~\ref{fig16}-\ref{fig17}.




\subsection{Sn isotope chain: evolution of the PDR strength}

\begin{figure}[b]
\includegraphics[scale=0.3]{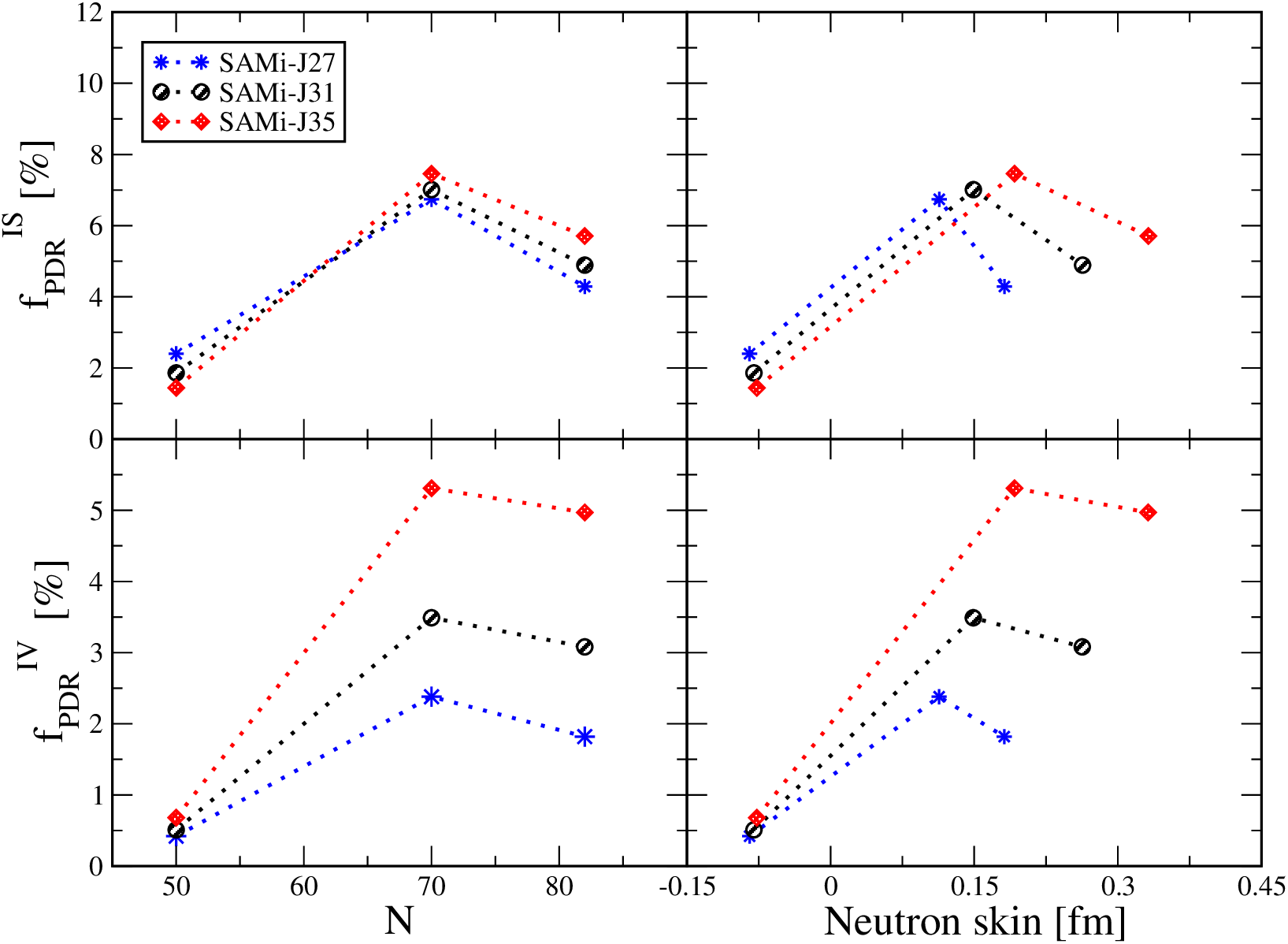}
\caption{(Color online) Fraction of the EWSR exhausted 
in the PDR region, for the IS (top) and IV (bottom) response, as a function of the neutron
number N and the neutron skin of the three Sn isotopes considered.}
\label{fig22} 
\end{figure}

In this section, we aim at assessing the evolution of the PDR strength when varying the N/Z ratio of nuclear systems. The analysis developed in recent works~\cite{ebaPRC2014} looks indeed mostly at the isospin asymmetry dependence of the percentage fraction of the Energy Weigthed Sum Rule (EWSR), $f_\textup{PDR}$, exhausted in the pygmy region of the IV dipole response (i.e. by the PDS). Here our goal is instead to establish a connection between the N/Z dependence of the IV dipole response and the concomitant behavior exhibited by the IS dipole strength, in view of the isoscalar-isovector mixing existing in neutron-rich systems and discussed above.
As already observed by Ebata {\it et al.}~\cite{ebaPRC2014}, despite the increase of the neutron skin thickness, the percentage fraction of the isovector EWSR $f_\textup{PDR}$ does not grow along the Sn isotope chain, when increasing the neutron number from $N = 70$ to $N = 82$. This result appears unexpected, considering the {relation discussed in several works} 
between the neutron skin thickness and 
the PDS in the IV dipole response~(see \cite{zhePRC2016} and Refs. therein). 
Although 
detailed shell structure effects can be invoked to solve this puzzle, we note that
a gateway to this uncommon behavior can be reached on the basis of the isoscalar-isovector mixing discussed so far. The missing fraction in $f_\textup{PDR}$ could be indeed attributed to the decrease observed in the isoscalar dipole strength, when Sn isotopes from $N = 70$ to $N = 82$ are considered. Although the fraction of the EWSR exhausted {in the low-energy region of} the IV dipole response {is expected to} increase for nuclei with a larger imbalance in neutron and proton numbers, the result is correlated also to the behavior of the IS response, which in turn reflects the evolution of the isoscalar density profile, when moving from $^{120}$Sn to $^{132}$Sn (see Fig.~\ref{fig19}). 

\begin{figure}[b]
\includegraphics[scale=0.3]{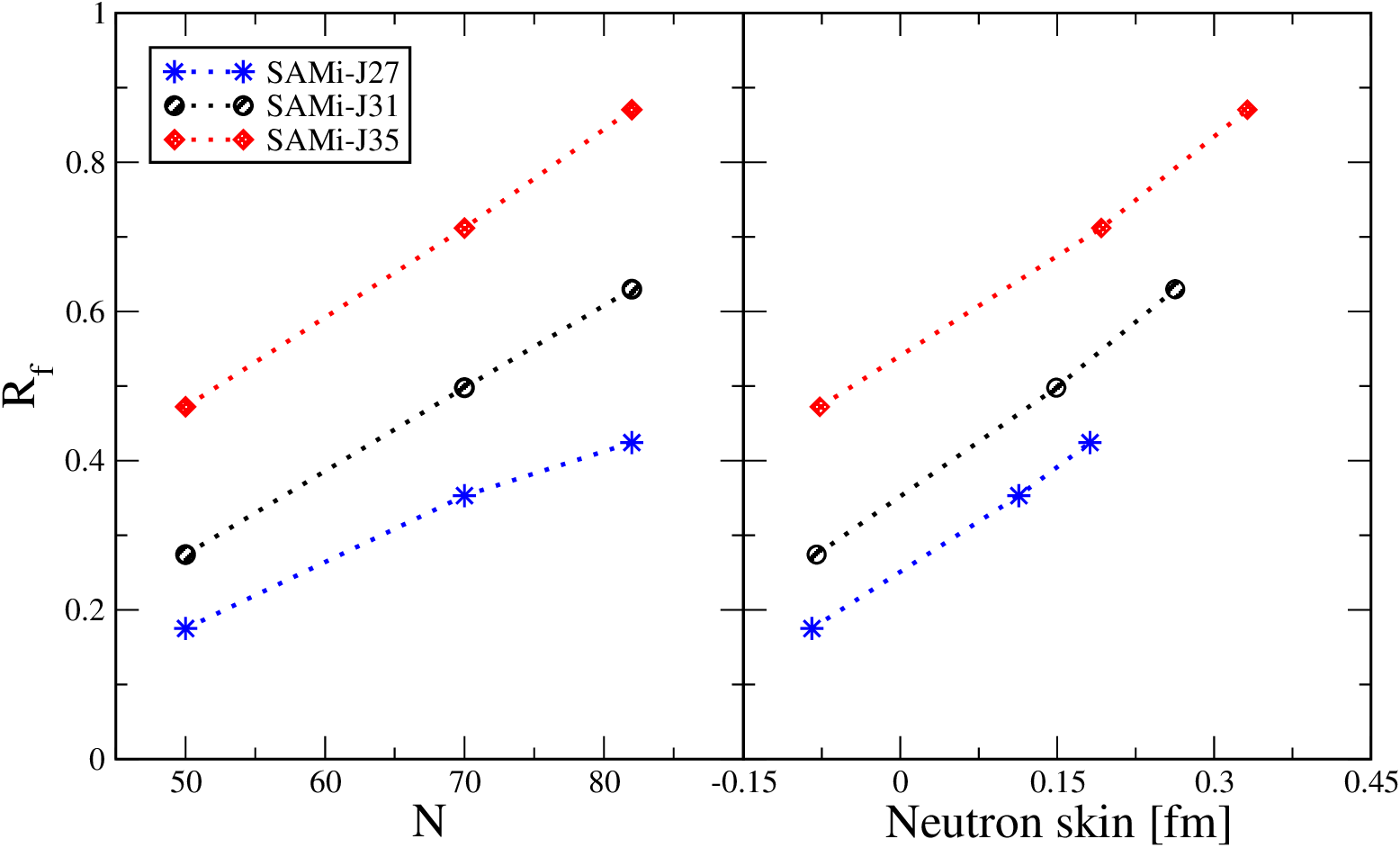}
\caption{(Color online) Ratio between fractions of EWSR exhausted in the PDR region of IV and IS response, as a function of the neutron
number N and the neutron skin of the three Sn isotopes considered. }
\label{fig23} 
\end{figure}

Fig.~\ref{fig22} represents the trend of the fraction of the EWSR exhausted 
in the PDR region {(below 10.5 MeV for $^{120}$Sn and $^{132}$Sn and below 11.3 MeV for $^{100}$Sn)}
for the IS and IV response, as a function of the neutron
number (left) and the neutron skin (right) of the three Sn isotopes.
To better isolate the PDR contribution in the dipole response, 
a width $\gamma$ = 0.5 MeV has been used in the cut-off function of Eq.~(\ref{eq:smooth}).  
One can see that our calculations reproduce the trend discussed, for the 
IV response (bottom panels), in Ref.~\cite{ebaPRC2014}, with an increase of the EWSR fraction
up to N=70 and then a decrease, though the neutron skin thickness is larger
in nuclei with larger N.  However, in Fig.~\ref{fig22} (top panels) 
one can see that the
same trend is exhibited also by the IS strength, owing to the prominence of 
the IS PDR strength in $^{120}$Sn, as discussed above. It is also interesting 
to notice that the IS $f_{\textup{PDR}}$ does not depend much on the effective 
interaction considered.

Then, to normalize the effect of the IV mixing to the strength of the mode considered,
that is mostly isoscalar,
we consider the ratio, $R_f$, between the EWSR fractions obtained in the IV and IS response. 
This quantity is shown in Fig.~\ref{fig23}, where
a nearly linear increase versus neutron number and neutron skin is 
now nicely observed.   
Thus we conclude that, according to the models employed in our study, the evolution of 
the PDR strength along an isotopic chain is not simply related to the neutron skin
thickness. Other ingredients may enter into game as well; indeed a deeper insight into the PDS
is got by looking, in parallel, at the corresponding IS strength. 
Moreover, one can notice that different parametrizations lead to
different results for the  $R_f$ ratio,  even when they predict close values of the neutron skin thickness (see the right panel of Fig.~\ref{fig23}). 
This indicates that, for a given asymmetric system, mixing effects are enhanced for effective interactions with larger symmetry energy slope L, as observed
in nuclear matter calculations \cite{barPR2005}.    

\section{Conclusions}

In this article, we have explored the features of the 
small-amplitude dipole response in selected nuclei 
within three approaches: TDHF, its zero-amplitude limit (RPA) and its
semi-classical limit (Vlasov).
 
As far as TDHF and RPA calculations are concerned, 
a detailed comparison of the dipole IS and IV strengths, and of the
transition densities of the main excitation modes, is presented here for 
the first time, showing a good agreement between the results of the 
two approaches. 

The comparison between quantal and semi-classical calculations 
has evidenced the importance of shell effects and quantal intrinsic gradient 
terms in shaping isoscalar and isovector density profiles of the 
ground state configuration. In particular, HF calculations
are generally associated with smoother isoscalar density profiles{, with respect to the ones deduced within the TF approximation}. 
Whereas the quantal IV dipole strength
is quite well reproduced by Vlasov calculations, significant differences
are observed in the low-energy domain of the IS response, concerning the
energy and the relative weight of the different peaks.   
Considering that this region is populated by surface excitation modes, 
this observation can be reconducted to the different density profiles 
and the different treatment of surface effects and gradient terms in 
quantal and semi-classical approaches.
{
Moreover, shell effects can affect significantly the details of the low-lying states, especially as far as the PDS is 
concerned.}


The low-energy region of the dipole response has been investigated in deeper detail. 
A thorough analysis of the associated transition densities allows one to characterize the different
modes in terms of IS/IV mixing and volume/surface components.
In particular, we observe that the lowest energy peak, in the PDR region,
corresponds to  
an isoscalar-like surface mode, of larger strength in nuclei with a more
diffuse surface.   The corresponding IV contribution, i.e. the PDS, 
originates from mixing effects and increases with the slope L 
of the symmetry energy. 
This trend stems from the fact that, as pointed out in several previous
investigations (see for instance \cite{carPRC2010}), a larger L is associated with a neutron enrichment of 
the surface region; moreover a larger derivative of the symmetry energy
also induces stronger {IS/IV} coupling effects, as indicated 
by nuclear matter calculations \cite{barPR2005}. 
Then one can argue that, for a considered system, 
the shape of the low-energy IS response is influenced by the
isoscalar density profile, whereas the strength of the corresponding 
IV counterpart is connected, among other effects, to the
surface neutron content, i.e. to the neutron skin thickness. 
These conclusions hold also in the semi-classical limit. 
This link to ground state properties may also
help to better understand the impact of relevant terms of the nuclear
effective interaction (and nuclear EoS), such as surface gradient terms, 
compressibility and symmetry energy, on the dipole response features. 

Looking at the dipole response of Sn isotopes, we observe 
a similar IS strength, in the PDR region,  in nuclei with similar density profile (once
rescaled by the nuclear radius), such as $^{100}$Sn  and $^{132}$Sn, whereas a larger
strength appears in the case of $^{120}$Sn, which exhibits a 
{more diffuse surface}.  The corresponding IV projection follows a similar behavior, being larger in the  $^{120}$Sn case, in spite of the thicker
neutron-skin of  $^{132}$Sn.  An increasing trend with the neutron skin thickness
can be recovered if one considers the ratio between the IV and IS EWSR 
fractions exhausted by the PDR region.


\section{Acknowledgments}
{Funding from the European Union's Horizon 2020 research   and   innovation program under Grant No. 654002 is acknowledged.} 


\end{document}